\documentclass[aps,prb,10pt,twocolumn]{revtex4-2}

\usepackage{graphicx}
\usepackage{dcolumn}
\usepackage{color}
\usepackage{amsmath}
\usepackage{xfrac}
\usepackage{amssymb}
\usepackage{bm,dsfont}
\usepackage{multirow}
\usepackage{xcolor}
\usepackage[pdftex,colorlinks=true,linkcolor=blue,citecolor=magenta,filecolor=blue]{hyperref}
\usepackage{comment}

\newcommand{\ff}[1]{{\boldsymbol #1}}
\newcommand{\ca}[1]{{\cal #1}}
\newcommand{\bi}{\begin{itemize}}
\newcommand{\ei}{\end{itemize}}
\newcommand{\be}{\begin{equation}}
\newcommand{\ee}{\end{equation}}
\newcommand{\ba}{\begin{eqnarray}}
\newcommand{\ea}{\end{eqnarray}}

\newcommand{\refeq}[1]{Eq.\ (\ref{eq:#1})}
\newcommand{\labeq}[1]{\label{eq:#1}}

\newcommand{\hs}{\hat{\ff s}}
\newcommand{\hS}{\hat{\ff S}}

\newcommand{\cb}{C^{\sf (B)}}

\newcommand{\ck}{C^{\sf (k)}}
\newcommand{\cs}{C^{\sf (S)}}
\newcommand{\scs}{c^{\sf (S)}}

\newcommand{\titlepaper}{Topology and Quantum-Spin-Classical-Spin Crossover of the Gapped Kondo Effect}

\begin{document}

\title{\titlepaper}

\author{David Kr\"uger}
\author{Michael Potthoff}
\affiliation{I. Institute of Theoretical Physics, Department of Physics, University of Hamburg, Notkestra{\ss}e 9-11, 22607 Hamburg, Germany}

\date{\today}

\begin{abstract}
The gapped Kondo effect describes the screening of an $S=\sfrac12$ impurity spin locally coupled via an antiferromagnetic exchange interaction to a conduction-electron system exhibiting a finite hard gap.
Using a combination of a Lanczos transformation and a self-consistent configuration-interaction scheme, we numerically investigate the local phase diagram. 
Furthermore, we show that the different phases can be characterized by several topological invariants: 
the conventional momentum-space Chern number of the underlying two-dimensional host system, corresponding to a Chern insulator; the B-space Chern number, defined by coupling the impurity spin to a fictitious local magnetic field $\ff B$, in the limit $B \to 0$; and the S-space Chern number, defined for a classical impurity spin, i.e., a vector of fixed length.
The classical-spin limit is obtained for $B \to \infty$. 
By varying the field strength, we can therefore continuously interpolate between quantum-impurity-spin and classical-impurity-spin Hamiltonians and investigate whether the corresponding phase diagrams are likewise continuously connected.
The gapped underscreened Kondo effect is studied for an impurity spin $S>\sfrac{1}{2}$ as well as in the classical-spin limit approached via $B \to \infty$ or $S \to \infty$. 
Different variants of scattering theory are employed to interpret the resulting phases.
Finally, the gapped two-channel overscreened Kondo effect, realized by coupling a quantum impurity spin equally to the local electron spins of both orbitals within a unit cell, is shown to be characterized by spontaneous particle-hole symmetry breaking. This leads to a highly nontrivial quantum–classical phase diagram.
\end{abstract}

\maketitle

\section{Introduction}
\label{sec:intro}

The Kondo effect \cite{Kon64,Hew93} is a central many-body paradigm in condensed-matter physics. 
A spin-$\sfrac12$ magnetic impurity is screened and a many-body singlet is formed at temperatures $T$ below the Kondo temperature $T_{\rm K} \sim e^{-1/\rho J}$, where $\rho$ is the density of states and $J$ is the strength of the antiferromagnetic local exchange coupling.
If the impurity is coupled to an insulator with a hard gap $\Delta>0$, a remnant of the Kondo effect remains.
For $T=0$, the singlet state forms at a finite critical coupling $J_{\rm c}$, roughly determined by $T_{\rm K}(J_{\rm c}) \sim \Delta$. 
The transition is expected to result from a level crossing of subgap states and is therefore discontinuous. 
This is different from the standard Kondo effect and also from the pseudogap Kondo effect \cite{WF90,BPH97,FV13}, where $\rho(\omega) \sim |\omega|^{r}$ with some finite $r>0$. 

The gapped Kondo effect, $\Delta > 0$, has been studied in different contexts, such as Kondo self-screening \cite{LLSN13,KAN14} or
BCS-type gapped superconductors with one or two quantum-spin impurities \cite{ZLL+10,ZBP11,YMW+14,PP24}. 
Another variant is the Kondo box, in which one \cite{TKvD99,Sch01,Sch02,SA02,CB02,FRF03,RM06,HKM06,BBH10} or more impurity spins \cite{SGP12,SHP14,SHPM15} are coupled to a nanosystem with fixed total particle number $N$ and finite-size gap $\delta$. 
One finds a pronounced fermion parity effect and, in the case of odd parity and $T=0$, a crossover to a Kondo-screened state at $J_{\rm K} \sim \delta$.

Treating magnetic impurities as {\em classical} spins, i.e., vectors of fixed length, reduces the many-body problem to a noninteracting one.
While this simplification is questionable \cite{SSFvO22}, it is widely used, e.g., for magnetic adatoms on superconductors \cite{SBS+20,KCZ+21,vOF21,FBBO21,DHLK21,MCL21,LRR+22,MGDW22,SMFvO22,ZZN+23}, where subgap energy bands due to hybridization of Yu-Shiba-Rusinov bound states \cite{Yu65,Shi68,Rus69} can host Majorana modes. 
Impurities in the bulk of a topological insulator \cite{HK10,QZ11,RSFL10} represent zero-dimensional defects. 
Their classification within the bulk-defect correspondence \cite{RSFL10,TK10,CTSR16} relies on noninteracting systems.

Here, we study the gapped Kondo effect for a quantum-spin impurity from the new perspective of {\em local topology}.
We first demonstrate that the transition can be characterized as a topological transition between two states with different values of a Chern number $\cb$. 
This invariant characterizes the bundle of ground states over a two-dimensional (2D) manifold (``B space'') given by the possible directions $\ff n \equiv \ff B / B$ of a weak fictitious external magnetic field $\ff B = B \ff n$ that only couples to the impurity spin $\hS$.
The Chern number $\cb$ is quantized, $\cb \in \mathbb{Z}$, with $\cb=0$ in the screened and $\cb \ne 0$ in the unscreened state, and represents a topological invariant that is spatially {\em local} but global in B space, which must be clearly distinguished from the conventional (``k-space'') Chern number $\ck$ that refers to the Brillouin zone (BZ) in a 2D system.

We furthermore show that the {\em strength} $B$ of the fictitious field can be used to continuously deform the quantum-spin into the classical-spin system.
For $B \to \infty$, the quantum spin $\hS$ behaves exactly like a classical magnetic moment $\ff S$, and thus the problem becomes uncorrelated. 
It turns out that the quantum-classical deformation of the gapped Kondo effect has a counterpart on the classical side, which in fact has recently been recognized as a local topological transition \cite{MFQ+24}. 
This transition is characterized by another local Chern number $\cs$, which is defined over the S-space of the directions of the classical spin at fixed length $|\ff S|$.

Analyzing the nontrivial topology carried by spatially local, closed 2D parameter manifolds, in addition to the nonlocal Brillouin zone, is also useful in more complicated settings. 
Here, we address quantum-spin impurities with spin quantum numbers $S > \sfrac12$, i.e., the underscreened gapped Kondo effect.
The approach can also be applied to systems with several impurity spins and even to gapped Kondo-{\em lattice} models.
Here, we demonstrate that the two-channel (overscreened) Kondo effect \cite{NB80}, for $\Delta>0$ \cite{SHPM15}, is accompanied by spontaneous particle-hole symmetry breaking below a critical strength of the local magnetic field.

We substantiate our conceptual approach by tracing local quantum phase transitions in a generic model system. 
For this purpose, we developed a method based on a Lanczos transformation \cite{BMF13,Hay80} combined with a self-consistent configuration-interaction scheme \cite{LCHH19}.
In the case of the underscreened gapped Kondo effect, a scattering-theory approach provides additional insight.

The paper is organized as follows:
Section \ref{sec:mod} introduces the generic model studied and gives a brief overview of the computational many-body approach. 
We present and interpret our results in Sec.\ \ref{sec:res}, starting with the phase diagram for a single $S=\sfrac12$ quantum-spin impurity in Sec.\ \ref{sec:qu} and a discussion of the implications of k-space and B-space topology in Sec.\ \ref{sec:top}.
In Sec.\ \ref{sec:cl} we use the local magnetic field to study the classical-spin phase diagram and discuss the continuous deformation between the phase diagrams for classical and quantum impurity spins.
The underscreened gapped Kondo effect, mainly for $S=1$, is addressed in Sec.\ \ref{sec:under}, and a systematic study of the $S$ dependence in the k-space topologically trivial and nontrivial regime is presented in Sec.\ \ref{sec:highs}. 
Section \ref{sec:scatt} presents insights obtained from exact and approximate scattering-theory approaches for a classical or quantum-spin impurity, respectively.
The phase diagram of the overscreened gapped Kondo effect is presented in Sec.\ \ref{sec:over}. 
The quantum-classical deformation in the overscreened case and a spontaneous symmetry breaking ruling the corresponding phase diagram is discussed in Sec.\ \ref{sec:ssb}. 
Concluding remarks are given in Sec.\ \ref{sec:con}.

\section{Generic model and computational approach}
\label{sec:mod}

We first consider an $S=\sfrac12$ spin $\hS$ coupled to the spinful Qi-Wu-Zhang (QWZ) model \cite{QWZ06}, i.e., a prototypical two-dimensional insulator with nontrivial k-space topological phases:
\be
H = H_{0} + H_{J}
= 
\sum_{\ff k \alpha \alpha' \sigma} \epsilon_{\alpha\alpha'}(\ff k) c_{\ff k \alpha\sigma}^{\dagger}c_{\ff k\alpha'\sigma}
+
J \hs \cdot \hS
\: .
\labeq{ham}
\ee
$\hS$ is antiferromagnetically coupled ($J>0$) to the local electron spin $\hs$ at the impurity site $i=i_{0}$. 
The spinful QWZ model $H_{0} = \sum_{\sigma=\uparrow, \downarrow} H_{0\sigma}$ has two orbitals $\alpha = \mbox{A},\mbox{B}$ per site.
The electron spin at site $i_{0}$ and in orbital A is given by $\hs \equiv \hs_{i_{0}A} = \sfrac12 \sum_{\sigma\sigma'} c_{i_{0}A\sigma}^{\dagger} \ff \tau_{\sigma\sigma'} c_{i_{0}A\sigma'}$, where $\ff \tau = (\tau_{x}, \tau_{y}, \tau_{z})$ are the Pauli matrices and $c_{i\alpha\sigma} = L^{-1/2} \sum_{\ff k} e^{i\ff k \ff R_{i}} c_{\ff k\alpha \sigma}$ are the local annihilator operators, and where $\ff R_{i}$ are the position vectors to the $L$ sites of the square lattice.
For brevity, we write $\hs \equiv \hs_{i_{0}A}$.
The $2 \times 2$ Bloch matrix in \refeq{ham} is
\ba
\ff \epsilon(\ff k)
  &=&
  \left[
  m + t \cos (k_{x}) + t \cos (k_{y}) 
  \right]
  \ff \tau_{z}
\nonumber \\
  &+&
  t \sin(k_{x}) \ff \tau_{x} + t \sin(k_{y}) \ff \tau_{y}
\: . 
\labeq{bloch}
\ea
Here, the Pauli matrices define the orbital structure of the system, and wave vectors $\ff k$ run over the first Brillouin zone (BZ).
The energy scale is given by the nearest-neighbor hopping $t=1$.

The resulting dispersions of the two bands are:
\be
\epsilon_{\pm}(\ff k) = \pm \left[ t^{2} \sum_{\nu=x,y} \sin^{2}k_{\nu} + (m+ t \sum_{\nu=x,y} \cos k_{\nu})^{2} \right]^{1/2}
\labeq{disp}
\: .
\ee
The band gap is $\Delta = 2 \times \mbox{min}_{\ff k} | \epsilon_{\pm}(\ff k) |$. 
With $t=1$, we have $\Delta = 2m$ for $0 \le m \le 1$, 
and $\Delta = 4-2m$ for $1 \le m \le 2$, 
and $\Delta = -4+2m$ for $2 \le m$.
At the critical points $m=-2,0,2$, there are semimetal phases.
The band gap closes at critical high-symmetry points in the BZ, namely at $\ff k_{\rm c} = (0,0)$, at $\ff k_{\rm c}=(0,\pi),(\pi,0)$, and at $\ff k_{\rm c}= (\pi,\pi)$, respectively.
For $-2 < m < 0$ (Chern number $\ck=+2$) and $0 < m < 2$ ($\ck=- 2$), the QWZ model is a Chern insulator in symmetry class A of the Altland-Zirnbauer scheme \cite{AZ97,RSFL10}. 
We have $\ck=\pm 2$ rather than $\ck =\pm 1$ due to the additional spin degree of freedom. 
Finally, for $m<-2$ and $m>2$, the system is a conventional band insulator. 

To compute various observables, such as the ground-state and excited-state energies $E_{0}, E_{1}, ...$, the local spin correlations $\chi_{\rm loc} = \langle \hs \cdot \hS \rangle$, and the local single-particle Green's function $G_{i\alpha\sigma}(\omega) = \langle\langle c_{i\alpha\sigma} ; c^{\dagger}_{i\alpha\sigma} \rangle \rangle$ at sites $i$ near $i_{0}$, we employ a two-step numerical technique. 

In the first step, we employ the Lanczos algorithm to construct a one-particle Krylov space $\ca K$ generated by the seed orbital $| i=i_{0}, A, \sigma \rangle$. 
This Lanczos transformation \cite{BMF13,Hay80} exactly maps the problem onto a semi-infinite chain geometry with $| i_{0}, A, \sigma \rangle$ represented as the first site. 
We formally consider a system in the thermodynamic limit $L \to \infty$, but work with a large but finite Krylov-space dimension $d = \ca O (100)$. 
Provided that the electronic-structure gap exceeds the finite-size gaps, the results no longer depend on the choice of $d$ as we have checked regularly.
Importantly, since the transformation leaves $| i_{0}, A, \sigma \rangle$ invariant, the Kondo coupling is still treated exactly.

In the second step, the resulting Kondo problem in the chain geometry is tackled with a restricted active-space configuration-interaction (RASCI) scheme based on Ref.\ \cite{LCHH19}. 
Operationally, we iteratively compute the single-particle reduced density matrix $\rho_{mm'} = \langle c_{m'}^{\dagger} c_{m} \rangle$ with $m \equiv (i,\alpha,\sigma)$. 
We start by expressing the Hamiltonian in the basis of natural orbitals obtained from $\ff \rho$ at $J=0$, excluding the impurity and thus leaving the interaction term invariant.
Diagonalization of the Hamiltonian in a truncated RASCI basis yields the ground state and hence an updated $\ff \rho$.
This is iterated until self-consistency.
Typical active-space dimensions range from 12 to 14.
See Appendix \ref{sec:num} for details of the theory. 

To study the system at half filling, we fix the chemical potential at $\mu=0$. 
The corresponding ground state is found in the sector with particle number $N=N_{0}$ that minimizes the energy $E_{0}(N)$.
Changes in the particle number due to level crossings are characterized by $\Delta N(J) = N_{0}(J) - N_{0}(J=0)$, where the particle-number change is measured relative to the $J=0$ system. 
Calculations have been performed in sectors at and close to $N_{0}(J=0)$, with $N_{0} = 50$ -- $100$.

\section{Results}
\label{sec:res}

\subsection{Phase diagram for a quantum-spin impurity}
\label{sec:qu}

\begin{figure}[t] 
\centering
\includegraphics[width = 0.6\linewidth]{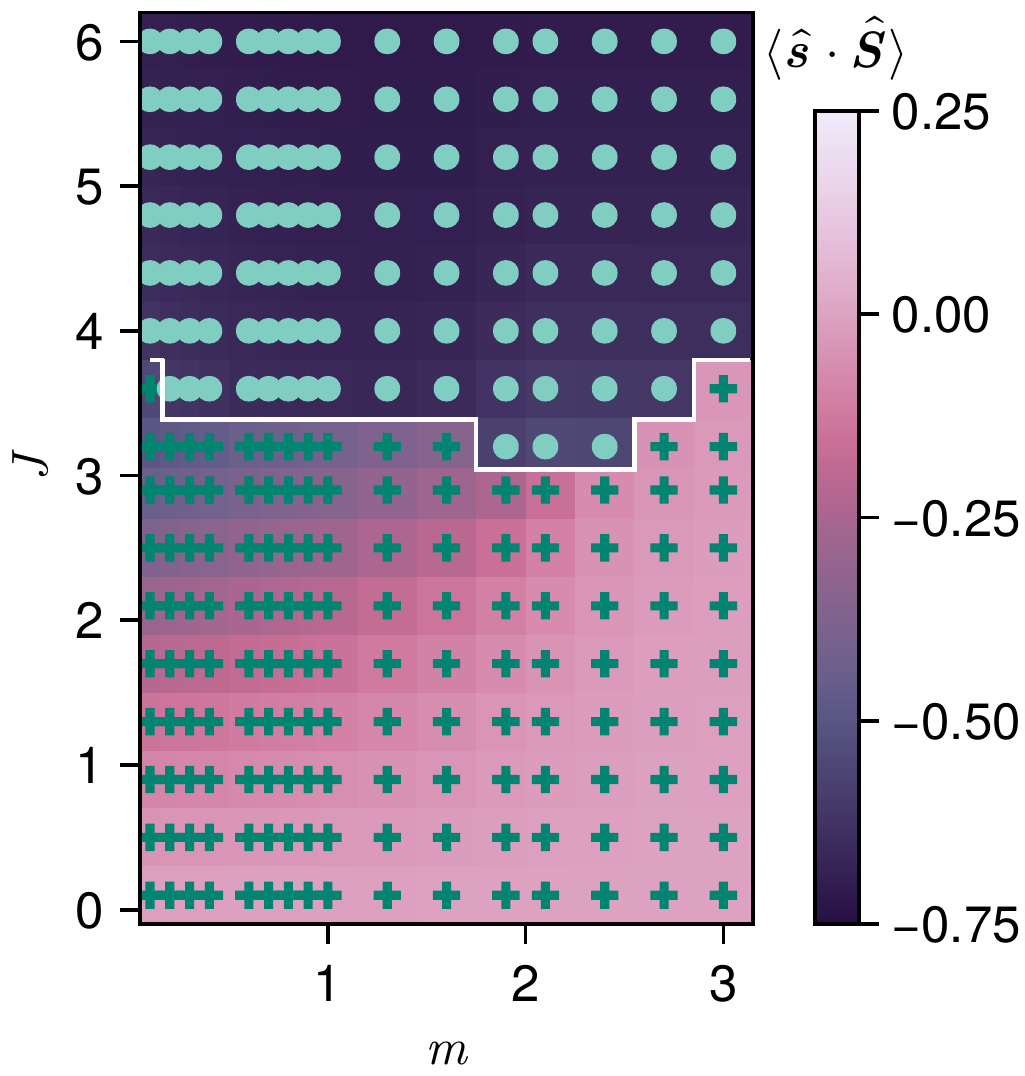}
\caption{
$m$-$J$ ground-state phase diagram. 
{\em White line:} discontinuous phase transition.
{\em Background color code:} local Kondo correlation $\langle \hs \cdot \hS \rangle \equiv \langle \hs_{i_{0} A} \cdot \hS \rangle$. 
{\em Light green symbol color:} $\Delta N = +1$, {\em dark green:} $\Delta N =0$. 
{\em Symbol form} indicates the ground-state degeneracy, doublet: {\em cross}, singlet: {\em circle}.
Calculations for $N = 50 + \Delta N$.
}
\label{fig:pd1}
\end{figure}

For a gapped electron system, the screening of the impurity spin is expected to set in at a {\em finite} Kondo coupling $J_{\rm K}$. 
In fact, we find a critical line $J_{\rm K}(m)>0$ for all $m>0$, which marks the boundary between the unscreened and the screened state, as shown in Fig.\ \ref{fig:pd1}.
Note that the phase diagram is symmetric under the sign change $m \to -m$ and $\Delta N \to-\Delta N$. 
For all $J < J_{\rm K}(m)$, including $J=0$, where the impurity spin $\hS$ is decoupled, the ground state is a total-spin doublet. 
With increasing $J$, the system develops weakly antiferromagnetic local spin correlations, $\langle \hs \cdot \hS \rangle < 0$ (see background color), mediated by virtual fluctuations across the gap on the scale $\sim t^{2}/\Delta$. 
An orbital polarization $\langle n_{i_{0}B} \rangle - \langle n_{i_{0A}} \rangle > 0$ induced by a finite mass parameter $m>0$ counteracts this effect, and for large $m$ we get $\langle \hs \cdot \hS \rangle \approx 0$ even for stronger $J$, but $J < J_{\rm K}(m)$.
On the other hand, in the topologically nontrivial regime $0<m<2$, screening via virtual fluctuations is enhanced due to the enhanced A character of the valence band near the edge, such that the impurity spin can access both sides of the gap.
This manifests itself in the considerably stronger correlation $|\langle \hs \cdot \hS \rangle|$ 
in this $m$ range and for $J < J_{\rm K}(m)$ (see also Appendix \ref{sec:cross}).

At $J = J_{\rm K}(m)$, we find that $\langle\hs \cdot \hS\rangle$ discontinuously jumps to a larger negative value.
This indicates the formation of a spatially extended Kondo singlet. 
As $J$ increases further, the local Kondo correlation quickly approaches $\langle \hs \cdot \hS \rangle = - \sfrac34$, i.e., the $J\to \infty$ value, where $\hS$ and the local spin $\hs$ of the electron system form a completely local Kondo singlet (LKS) at $i_{0}$.
At $J= J_{\rm K}(m)$, the A-orbital occupation at $i_{0}$ jumps from values $\langle n_{i_{0}A\sigma} \rangle < 0.5$, depending on $m$, to almost the $J\to \infty$ value $\langle n_{i_{0}A\sigma} \rangle = 0.5$.
The ground state for $J > J_{\rm K}(m)$ is a total-spin singlet in the sector with an additional electron, $\Delta N = +1$, as compared to $J < J_{\rm K}(m)$.

\begin{figure}[t] 
\centering
\includegraphics[width = 0.9\linewidth]{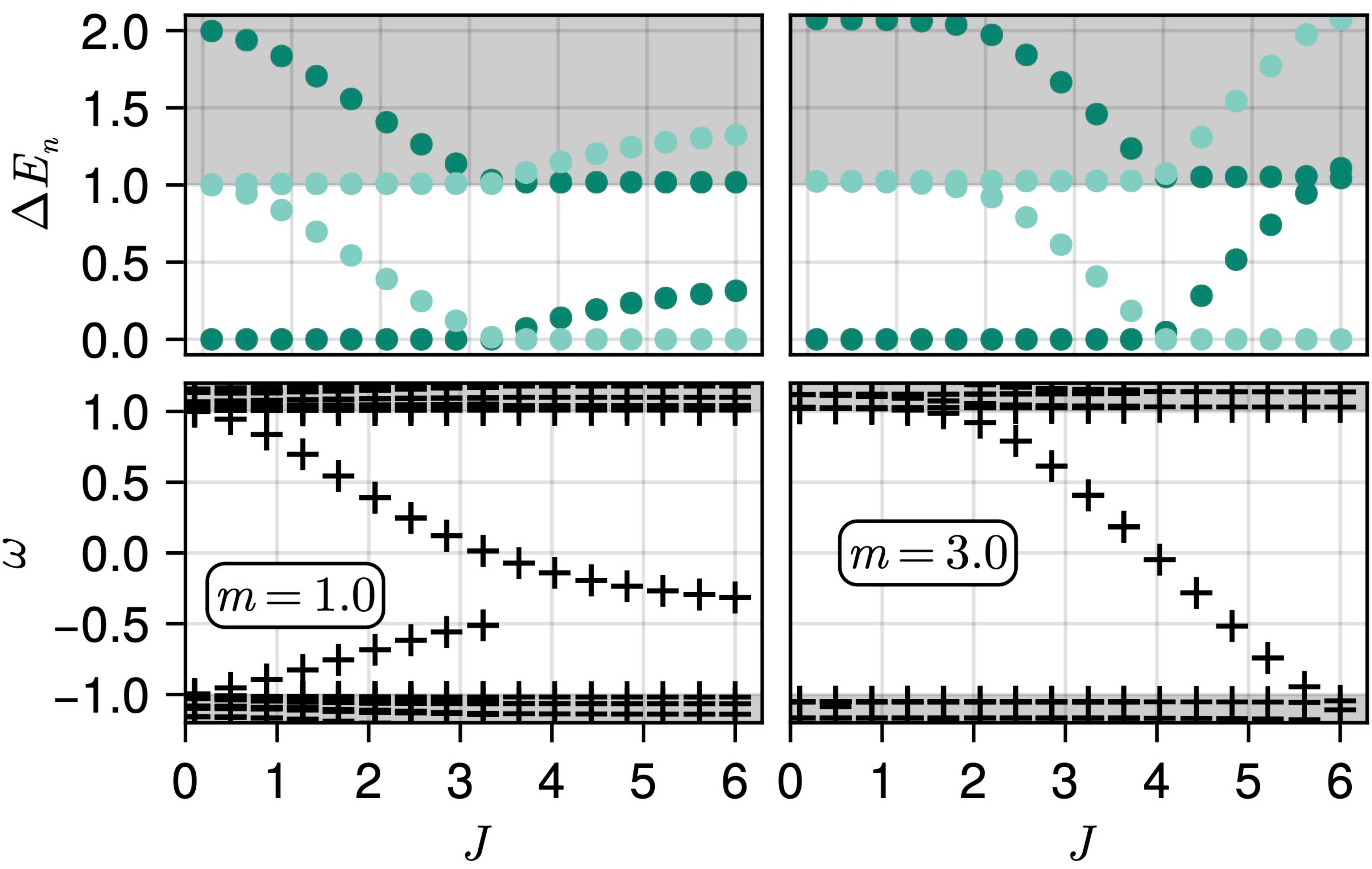}
\caption{
{\em Top:} $J$-dependent difference between the energies of the lowest excited states and the ground state, $\Delta E_{n} (J)\equiv E_{n}(J) - E_{0}(J)$.
{\em Dark green:} particle number $\Delta N=0$ relative to the $J=0$ ground state.
{\em Light green:} $\Delta N=1$.
{\em Bottom:} $J$-spectral flow of the low-frequency poles of the single-electron Green's function $\langle \langle c ; c^{\dagger} \rangle \rangle$.
Note that the bulk-band gap is $\Delta = 2$ for both $m=1$ and $m=3$.
{\em Gray area:} bulk continuum of single-particle excitations.
}
\label{fig:flow1}
\end{figure}

The discontinuous nature of the transition to the Kondo-screened phase is demonstrated in Fig.\ \ref{fig:flow1} (top panels). 
For $m=1$ (left, k-space topologically nontrivial with $\ck=-2$) 
and $m=3$ (right, $\ck=0$), we find a level crossing at $J_{\rm K} \approx 3.3$ and $J_{\rm K} \approx 3.9$, respectively.
The ground state switches between different sectors, from $\Delta N=0$ for weak $J$ to the $\Delta N=+1$ sector for strong $J$.
When coupling $\hS$ to a B orbital, we would have $\Delta N=-1$ for $m>0$.

Qualitatively, the particle-number change, the critical interaction, and the lowest excitation energies can be understood in a qualitative way as follows:
The ground state in the $\Delta N=0$ sector is approximately given by the completely filled valence band, i.e., by the Dirac sea state $| \mbox{sea} \rangle$ with energy $E^{(0)}_{0} = E_{\rm sea} = \sum_{\ff k} \epsilon_{-}(\ff k)$. 
Note that $| \mbox{sea} \rangle$ is not an eigenstate of $H_{J}$ and may serve as a good approximation for large gap $\Delta$ only.
The ground state in the $\Delta N = +1$ sector is approximately given by $| \mbox{sea} \rangle \otimes |\mbox{LKS}\rangle$, i.e., with an additional local Kondo singlet (LKS) formed by $\hs_{i_{0}A}$ and $\hS$. 
Compared to $E^{(0)}_{0}$, the LKS energy $E_{\rm LKS} = -\sfrac34 J$ is gained but the on-site energy $m>0$ must be paid, i.e., the ground-state energy in the $\Delta N=+1$ sector is $E_0^{(+1)} = E_{0}^{(0)} + m + E_{\rm LKS}$. 
Hence, the critical interaction $J_{\rm K}$ is obtained from the condition $m + E_{\rm LKS} = 0$. 
We thus obtain
\be
  J_{\rm K}(m) = \frac43 \, m
  \: .
\labeq{jk}
\ee
This rough estimate agrees almost perfectly with the phase boundary in Fig.\ \ref{fig:pd1} for $m \gtrsim 2$, also beyond $m=3$ (not shown), but breaks down in the k-space nontrivial regime $0 < m < 2$, cf.\ Ref.\ \cite{LLSN13}. 
It is instructive to compare with the phase diagram for a spin-$\sfrac12$ coupled to a trivial Semenov insulator, see Appendix \ref{sec:sem}, where an almost linear $m$ dependence is observed in the entire $m$ range.

The lowest excitation energies in the trivial phase at $m=3$, see Fig.\ \ref{fig:flow1} (top right), can be explained in a similar way: 
For $J > J_{\rm K}(m=3) \approx 4$, we (trivially) find for the first excitation energy $E_{0}^{(0)} - E_{0}^{(+1)} = - m - E_{\rm LKS} = - m + \sfrac34 J$. 
It increases linearly with $J$.
The second excitation energy, $E_{1}^{(0)} - E_{0}^{(+1)} = \Delta / 2$, is constant and corresponds to the excited state
$c_{\ff k_{\rm max}, -, \sigma} | \mbox{sea} \rangle \otimes | {\rm LKS} \rangle$ in the $\Delta N = 0$ sector.
Compared to the ground state $|\mbox{sea} \rangle \otimes | {\rm LKS} \rangle$, an electron from the valence-band maximum at $\ff k_{\rm max} = (\pi,\pi)$ is removed at the cost of half the single-particle gap $\Delta / 2 = - \epsilon_{-}(\ff k_{\max}) = 1$.
The third excitation energy, $E_{1}^{(+1)} - E_{0}^{(+1)} = -E_{\rm LKS} + \Delta / 2$, is linear in $J$ again and corresponds to the excited state
$c^{\dagger}_{\ff k_{\rm min}, + , \sigma} | \mbox{sea} \rangle$ in the $\Delta N = +1$ sector.
In contrast to the ground state $|\mbox{sea} \rangle \otimes | {\rm LKS} \rangle$, the additional electron at the conduction-band minimum is fully delocalized and cannot gain the local-Kondo-singlet energy.

Let us emphasize that all states discussed above should be regarded as rough approximations to the true many-body eigenstates. 
Particularly for the k-space topologically nontrivial phase for $m<2$, an improved understanding is required to explain why $J_{\rm K}$ is almost independent of $m$.

At $m=0$, the gap closes. Nevertheless, the $J=0$ local density of states on the A (and on the B) orbital vanishes, $\rho_{A}(\omega)=\rho_{B}(\omega) \propto \omega \to 0$ for $\omega \to 0$, i.e., here we have a pseudogap Kondo effect and thus expect a finite $J_{\rm K}$ \cite{FV13}.  
For small $m$, the critical line $J_{\rm K}(m)$ in Fig.\ \ref{fig:pd1} appears to increase with decreasing $m \to 0$. 
However, we also find that finite-system-size effects become increasingly important for $m \to 0$.
A tentative explanation for the slight upturn of $J_{\rm K}(m)$ is given in Sec.\ \ref{sec:scatt}.

\subsection{k-space and B-space topology}
\label{sec:top}

The transition also manifests itself in a pole of the single-electron Green's function crossing $\mu=0$, see lower panels of Fig.\ \ref{fig:flow1}.
Only in the nontrivial case $m=1$ ($\ck=-2$) does the pole stay {\em within} the gap, $\omega \rightarrow -0.5$ as $J \to \infty$. 
It corresponds to a single-electron excitation {\em bound} to the impurity but with zero weight on orbital A at site $i_{0}$. It is the remnant of the topologically protected mode at a one-dimensional boundary that has shrunk 
to the zero-dimensional boundary of the LKS, i.e., of the ``hole'' dynamically generated by suppressing hopping to and from A at $i_{0}$ for $J\to\infty$. 
This is the many-body or Kondo variant of the well-known bound states around zero-dimensional defects, which may serve as detectors for nontrivial k-space topology \cite{SLLS11,SRZB15,DFVR20,MFQ+24}.
Since the rest of the system without the hole is effectively uncorrelated for $J\to \infty$, the reasoning is the same as in the case of classical impurities and rigorous. 
Further discussion is given in Sec.\ \ref{sec:under} for a quantum-spin $S=1$ impurity.

The gapped Kondo effect is actually topologically enforced: 
We couple the impurity spin to a fictitious external field $\ff B= B \ff n$ with infinitesimally small field strength $B\searrow 0$
via $H \to H - \ff B \cdot \hS$. 
The set of possible field directions ${\sf B} \equiv \{ \ff n : \, |\ff n| = 1 \}$ forms a closed manifold ${\sf B} \cong S^{2}$, the B space, over which the quantized Chern number $\cb \in \mathbb{Z}$ can be computed as $\cb = (2\pi)^{-1} \int \ff F(\ff n) \cdot d\ff n $.
Here, $\ff F = \partial_{\ff B} \times \ff A$ is the Berry curvature, and $\ff A = i \langle 0, \ff B | \partial_{\ff B} | 0, \ff B \rangle$ is the Berry connection of the bundle of ground states $| 0, \ff B \rangle$ over {\sf B}.
Consider two extreme limits: 
For $J=0$, the impurity spin $\hS$ is decoupled from the electron system, such that the magnetic-monopole (Dirac) problem $H_{\rm eff} = - B \ff n \cdot \hS$ remains \cite{Dir31,Sim83,Ber84}, where we trivially have $\cb = -1$ \cite{Ber84}.
For $J \to \infty$, $\hS$ forms a LKS with $\hs$, such that a weak field has no effect, and hence $\cb=0$.
Therefore, to allow for a change of $\cb$, there must be a critical coupling $J_{\rm K}(m)$, {\em for all} $m>0$ ($m \ne 2$), at which the ground-state energy becomes degenerate. 

Here, this degeneracy is caused by a pole of $\ff G(\omega)$ crossing $\mu$. 
Recall our choice $\mu=0$. 
Since $\mu$ could have been chosen differently, but within the gap, the $J$-spectral flow of the poles must {\em fully} cross the band gap.
B-space topology thus {\em enforces} a gapless flow, as is seen in Fig.\ \ref{fig:flow1} (right {\em and} left).
This is accompanied by a change $\Delta N \ne 0$ of the particle number \cite{twop}.
 
\subsection{Classical-spin impurity and quantum-classical deformation}
\label{sec:cl}

For a classical impurity spin, i.e., replacing $\hS$ by a classical vector $\ff S$ of fixed length, $\hS \to \ff S = S \ff n$, a related topological analysis of the phase diagram is instructive.
To characterize the bundle of ground states over the closed two-dimensional space of spin configurations ${\sf S} \equiv \{ \ff n : \, |\ff n| = 1 \} \cong \mathbb{S}^{2}$, we use the corresponding S-space Chern number $C^{\sf (S)}$ \cite{MFQ+24}.
The latter is easily obtained in two limits:
At $J=0$, where the impurity spin is decoupled, we trivially have $C^{\sf(S)} = 0$.
Conversely, for $J \to \infty$, there is a dynamical decoupling of energy scales. 
At high energies of the order of $J$, the local physics is captured by the quantum-classical two-spin model $H_{\rm eff} = J \hs \cdot \ff S$. 
Again, this is the well-known Dirac monopole model \cite{Dir31,Ber84}, for which we immediately have $C^{\sf (S)}=1$, 
induced by the monopole at the singularity $J \ff S=0$ located inside $\mathbb{S}^{2}$. 
The remaining effective model $H^{\rm low}_{\rm eff}$ at low energies of $\ca O(t)$ is given by the non-interacting tight-binding model $H_{0}$, see \refeq{ham}, but with the A orbital at site $i_{0}$ removed. 
Hence, $H^{\rm low}_{\rm eff}$ does not contribute to the Chern number.

\begin{figure}[t] 
\centering
\includegraphics[width = 0.85\linewidth]{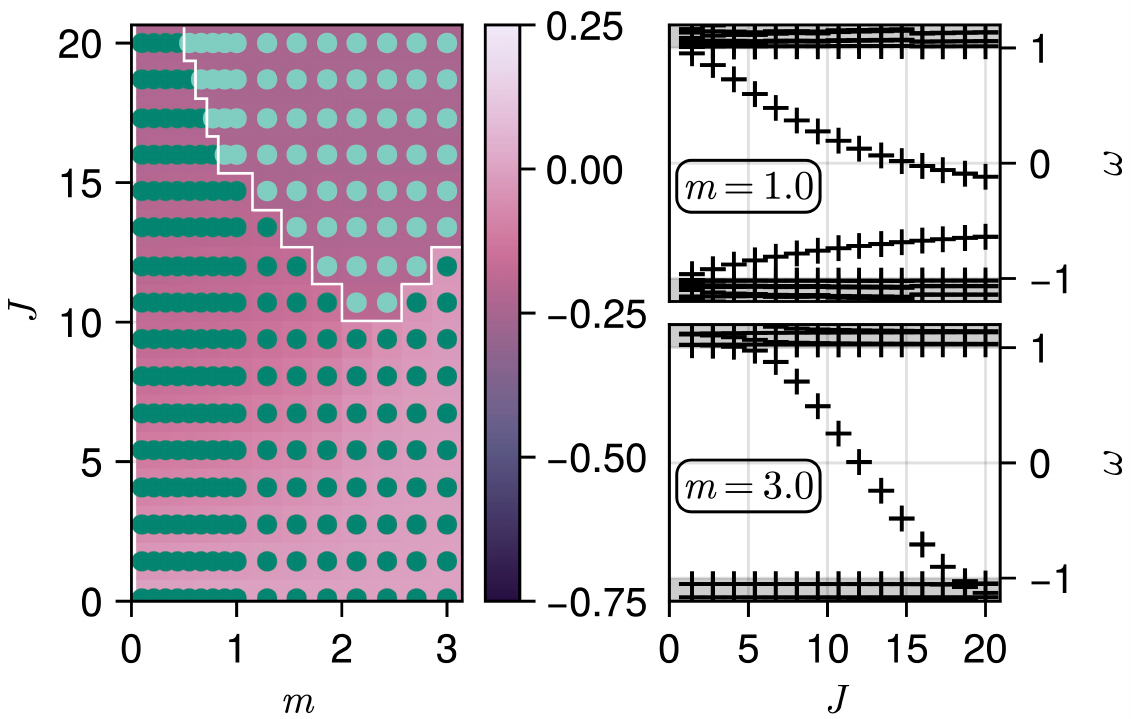}
\caption{
{\em Left:} $m$-$J$ phase diagram for a classical spin $\ff S$ of length $S=\sfrac12$.
{\em Background color code:} $\langle \hs \cdot \ff S \rangle = \langle \hs \rangle\cdot \ff S $. 
{\em Symbol color} indicates $\Delta N$, 
{\em green:} $\Delta N =0$, 
{\em light-blue:} $\Delta N = +1$.
Calculations for $N = 50 + \Delta N$.
{\em Right:} 
$J$-spectral flow of in-gap states for $m=1$ ({\em top}) and $m=3$ ({\em bottom}).
Bulk continuum: {\em gray background}.
}
\label{fig:pdclass}
\end{figure}

A change of $\cs$ requires the ground-state energy to be degenerate at some critical coupling $J_{\rm c}$, i.e., a single-electron excitation crossing $\mu$ and thus an associated particle-number change $\Delta N =  \pm 1$ at $J_{\rm c}$ \cite{twop}. 
The white line $J_{\rm c}(m)$ in Fig.\ \ref{fig:pdclass} (left) marks the phase boundary.
The $\cs=0$ phase with $\Delta N=0$ (green) smoothly evolves from the $J=0$ limit. 
The nontrivial $\cs=1$ phase with an additional electron $\Delta N=+1$ (light blue) connects to $J \to \infty$. 
We note that $J_{\rm c}(m)$ is at a minimum for $m \approx m_{\rm c} = 2$, which, for $J=0$, corresponds to the semimetal phase.
Comparing Figs.\ \ref{fig:pd1} and \ref{fig:pdclass} shows that $J_{\rm c}(m) \gg J_{\rm K}(m)$.
This is consistent with a modified energy balance, as is already seen in the atomic limit:
For a classical spin of length $S=\sfrac12$, the spin-coupling energy $-\frac14 J$ replaces the quantum-singlet energy $-\sfrac34 J$, and results in $J_{\rm c}(m) = 4m \gg \sfrac43 \, m = J_{\rm K}(m)$.

Importantly, both phase diagrams are qualitatively identical. 
The same holds for the $J$-spectral flow of the in-gap single-electron states, see Fig.\ \ref{fig:pdclass} (right). 
For the same reasons as in the quantum-spin case, (i) the flow must traverse the gap completely, and (ii) only in the k-space nontrivial case ($m=1$) does a state persist inside the gap for $J \to \infty$ as a remnant of the (k-space) topologically protected one-dimensional boundary mode.

Unlike the quantum-spin case, the spin-SU(2) symmetry of the system at $J=0$ is broken and states can be labelled by $\sigma=\uparrow, \downarrow$ referring to $\ff S$ as the spin-quantization axis.
For $J\to \infty$, due to the separation of energy scales, the mode residing in the gap must be spin degenerate and is, for finite but strong $J$, slightly spin-split.
For $m \to 0$, the numerical data are consistent with a diverging critical coupling $J_{\rm c}(m)$, contrary to the quantum-spin case.
This is confirmed by the exact scattering-theory approach in Sec.\ \ref{sec:scatt}.

The quantum-spin Hamiltonian can be deformed continuously into the classical-spin Hamiltonian. 
This can be achieved by increasing the strength of the fictitious external magnetic field that couples to the impurity spin as $H \to H - \ff B \cdot \hS$.
For $\ff B \to \infty$, all quantum fluctuations of $\hS$ are suppressed due to full spin polarization, such that $\hS$ becomes indistinguishable from a classical spin $\ff S$ of length $S = |\ff S| =\sfrac12$ that is fully aligned with $\ff B$.
This deformation raises the question of whether the corresponding phase diagrams are continuously deformable into each other as well.
Considering Chern numbers in the extreme limits $J=0$ and $J\to \infty$, there is no topological obstruction to such a deformation:
Recall that, in the quantum-spin case and for infinitesimally weak $B \to 0$, we have $\cb = -1$ for $J=0$ and 
$\cb = 0$ for $J\to\infty$, as discussed above, and that this enforces (at least) one transition at a coupling $J_{\rm K}(m)$.
On the other hand, when first taking the $B\to \infty$ limit, the impurity spin behaves classically and, as discussed above, we get $\cs=0$ for $J=0$ while $\cs =1$ for $J \to \infty$.
Hence, there must be (at least) one transition at some critical coupling $J_{\rm c}(m)$.
We note that quantum-spin and classical-spin cases are linked by the relation $\cb = \cs - 1$.

\begin{figure}[t] 
\centering
\includegraphics[width = 0.497\linewidth]{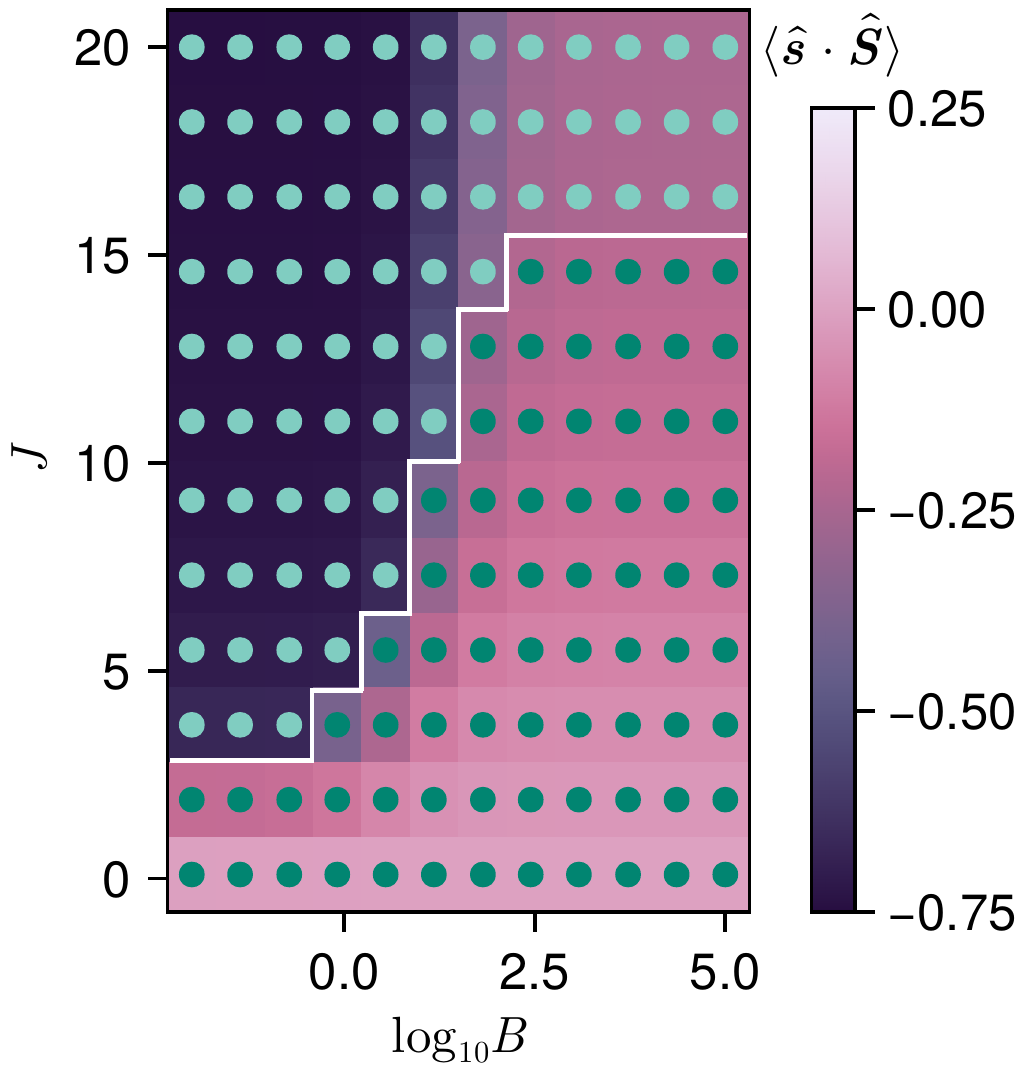}
\includegraphics[width = 0.49\linewidth]{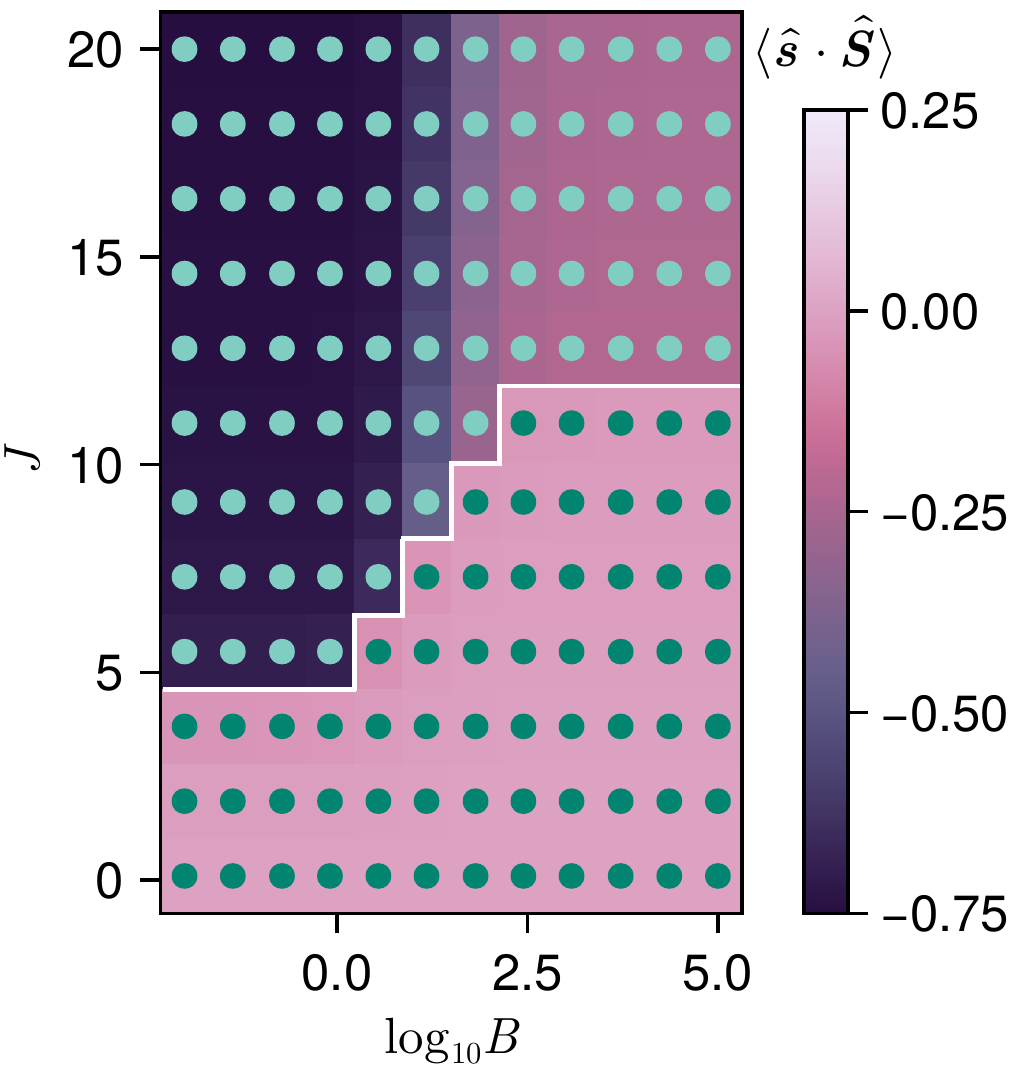}
\caption{
$B$-$J$ phase diagrams interpolating between quantum ($B\to 0$) and classical ($B\to \infty$) spin impurity models.
{\em Left:} $m=1$, {\em right:} $m=3$.
{\em Symbol color} indicates $\Delta N$, 
{\em green:} $\Delta N =0$, 
{\em light-blue:} $\Delta N = +1$.
Note the log scale.
}
\label{fig:bflow}
\end{figure}

Figure \ref{fig:bflow} shows the dependence of the critical coupling on $B$ (white lines) for two different mass parameters, $m=1$ (left) and $m=3$ (right), as obtained numerically.
In both cases, we see that the quantum-spin $\cb = -1$ phase for $J < J_{\rm K}(m)$ continuously connects with the classical-spin $\cs = 0$ phase for $J < J_{\rm c}(m)$ and, for strong $J$, the quantum $\cb = 0$ phase connects to the classical $\cs = 1$ phase.
Therefore, both the quantum-spin and the classical-spin phase diagrams are identical up to continuous deformations, though the quantitative differences are significant.

\subsection{Underscreened case}
\label{sec:under}

Considering an $S>\sfrac12$ rather than an $S=\sfrac12$ impurity spin $\hS$, we can repeat our considerations. 
For $J=0$ and for a given spin quantum number $S$, the B-space Chern number of the effective model $H_{\rm eff} = - \ff B \cdot \hS$ is now given by $\cb=-2S$ (see, e.g., Ref.\ \cite{Ber84} for the concrete calculation).
In the opposite limit $J \to \infty$, we first consider the interaction term $H_{J} = J \hs \cdot \hS$. 
The ground state is a $(2T+1)$-fold degenerate multiplet characterized by the spin quantum number $T$ of the total local spin $\ff T \equiv \ff s + \ff S$. 
For antiferromagnetic coupling $J>0$, we have $T=S-\sfrac12$. 
Coupling the resulting $2S$-fold degenerate multiplet perturbatively to the local magnetic field $B\searrow 0$, yields $\cb = -2S+1$.

For the classical-spin case, the S-space Chern numbers in the two limits are different as well: 
At $J=0$, we trivially have $\cs=0$, while for $J\to\infty$, the Chern number is obtained from the effective model $H_{\rm eff}= J \hs \cdot \ff S$ and given by $\cs = 1$, i.e., different from the $J=0$ value.
We conclude that for both the quantum and the classical cases, the transition is topologically enforced and there is no topological obstruction to deform the classical and the quantum phase diagrams into each other. 
We also note the relation $\cb = \cs - 2S$.

\begin{figure}[t] 
\centering
\includegraphics[width = 0.9\linewidth]{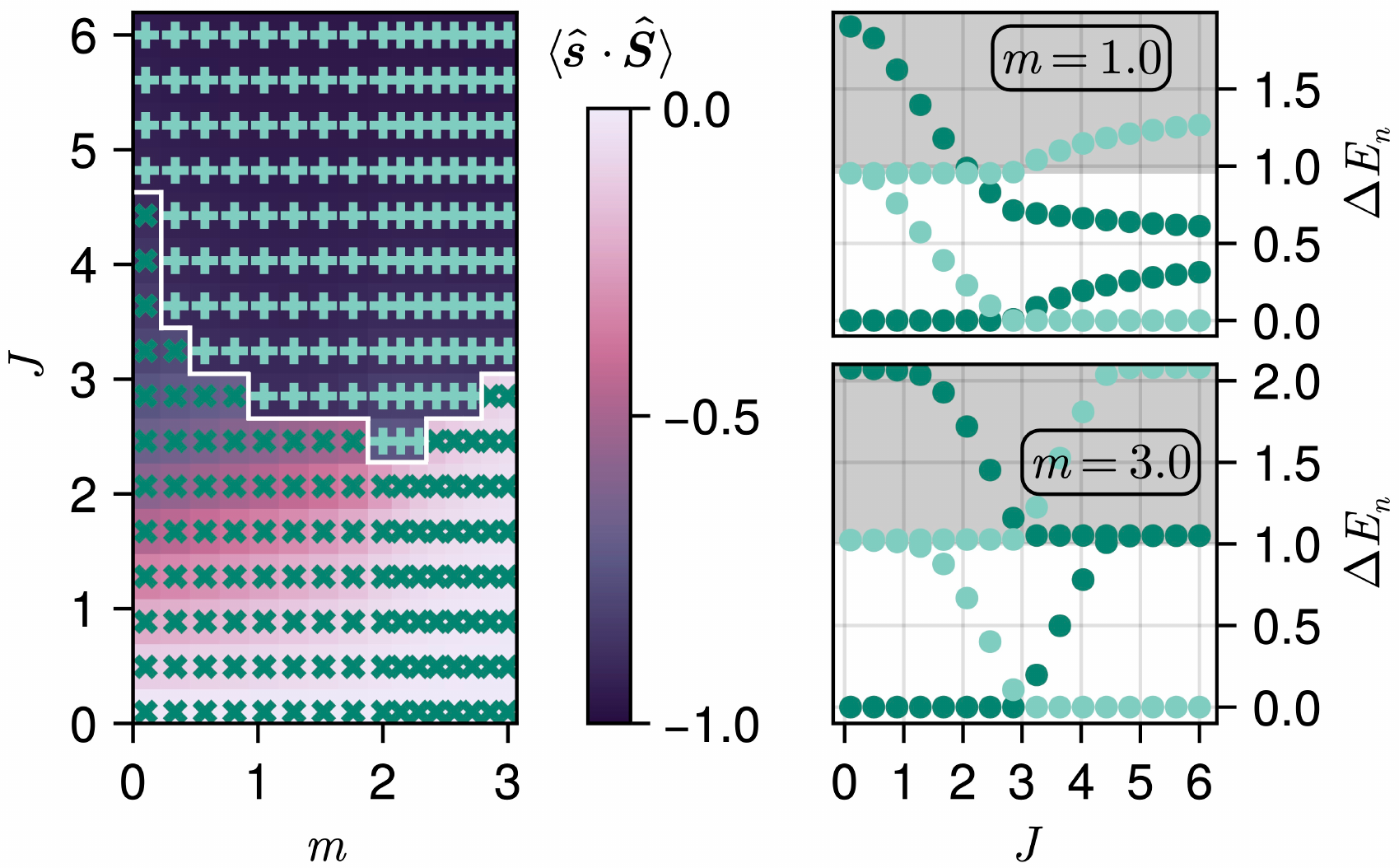}
\caption{
{\em Left:} 
$m$-$J$ phase diagram as obtained numerically for an $S=1$ quantum-spin impurity (underscreened case).
{\em White line:} discontinuous phase transition.
{\em Background color code:} local spin correlation $\langle \hs \cdot \hS \rangle$.
{\em Light blue symbol color:} $\Delta N = +1$, {\em green:} $\Delta N = 0$. 
{\em Symbol form:} ground-state degeneracy. Triplet: {\em cross}, doublet: {\em plus}. 
Calculations for $N = 50 + \Delta N$.
{\em Right:} 
$J$-dependent difference between the energies of the lowest excited states and the ground state, $\Delta E_{n} (J)\equiv E_{n}(J) - E_{0}(J)$ for $n=0,1,2,3$ and for $m=1$ ({\em upper panel}) and $m=3$ ({\em lower}).
{\em Dark green:} particle number $\Delta N=0$ relative to the $J=0$ ground state.
{\em Light green:} $\Delta N=1$.
{\em Gray area:} bulk continuum of single-particle excitations.
The bulk gap for electron-addition and for electron-removal excitations is $\Delta/2 = 1$ for both $m=1$ and $m=3$.
}
\label{fig:under-pd}
\end{figure}

The $m$-$J$ phase diagram for $S=1$ is shown in Fig.\ \ref{fig:under-pd}.
The discussion is in many respects analogous to the one for $S=\sfrac12$ and Fig.\ \ref{fig:pd1}.
However, for weak $J$ the ground state is a spin triplet with $\cb=-2$. 
It changes discontinuously at $J = J_{\rm K}(m)$ (white line) and is a spin doublet with $\cb=-1$ for $J > J_{\rm K}(m)$. 
No further screening of the $S=1$ impurity spin takes place such that it remains underscreened.
As compared to $S=\sfrac12$, the $m$-dependent phase boundary is very similar but with an overall slightly smaller $J_{\rm K}(m)$, except for small $m \to 0$, where $J_{\rm K}(m)$ increases earlier and stronger with decreasing $m$.
We also note that for the underscreened case and in the strong-$J$ limit, the local spin correlation approaches its minimal value, given by $\langle \hs \cdot \hS \rangle \to -1$.

In the regime $2 < m < \infty$ the critical coupling increases linearly with $m$. 
From Fig.\ \ref{fig:under-pd}, we quite precisely find $J_{\rm K}(m) \approx m$.
Note that this also holds for large $m \gg 2$ (not shown).
As for $S=\frac12$ (where $J_{\rm K} \approx \sfrac43 \, m$), this can be derived using the same qualitative reasoning [see Sec.\ \ref{sec:qu} below \refeq{jk}]:
For general $S$, the ground state in the sector $\Delta N=+1$ is approximately given by a tensor product of the filled Dirac sea $| \mbox{sea} \rangle$ with the $T=S-\sfrac12$ spin multiplet that results from the antiferromagnetic coupling of the impurity spin $S$ with the local spin $s=\sfrac12$ of the additional electron localized at $i_{0}$ on orbital A. 
Relative to the $\Delta N = 0$ ground state, its energy is $E_0^{(+1)} = E_{0}^{(0)} + m + E_{\rm ex}$. 
The mass term derives from the single A orbital occupancy. 
The exchange term is obtained by computing $J \langle \hs \cdot \hS \rangle = \sfrac12 J [ T (T+1) - S(S+1) - \frac34 ] = - \sfrac12 J (S+1)$ with $T=S-\sfrac12$.
The condition $m + E_{\rm ex} = 0$ then yields the critical coupling: 
\be
J_{\rm K}(m) = \frac{2m}{S+1}
\: .
\labeq{jks}
\ee
For $S=1$ we indeed find $J_{\rm K}(m) = m$ (and $J_{\rm K}(m) = \frac43 m$ for $S=\frac12$).

For $0 < m < 2$, \refeq{jks} fails to reproduce the $m$ dependence of $J_{\rm K}$, as in the $S=\sfrac12$ case (see Sec.\ \ref{sec:scatt} below for a refined analysis).
We also note that the simplified analysis runs into problems at $m=1$ for $J \gtrsim 4$, as this cannot address hybridization effects.

The $J$-dependent energy spectrum of the subgap states is shown in Fig.\ \ref{fig:under-pd} (right panels). 
For the k-space topologically trivial case at $m=3$ (lower panel), we see that a single state, with total local spin $T=1$, fully crosses the many-body gap as a function of $J$. 
This is reminiscent of the classical-spin case, where one can rigorously argue that S-space topology requires a full crossing.

The k-space topologically nontrivial case at $m=1$ (upper panel) exhibits a nontrivial subgap-state structure, which persists in the limit $J\to \infty$. 
This is easily analyzed, as we can profit from the separation of energy scales $J \gg t$. 
At strong $J > J_{\rm K}$, the overall ground state lies in the $\Delta N=+1$ sector and is a total-spin doublet.
For $J \to \infty$, it is given by a tensor product $| \mbox{sea} \rangle \otimes | \mbox{UKD}, m_{z} \rangle$ of the filled Dirac sea $| \mbox{sea} \rangle$ with the $T = S - \sfrac12 = \sfrac12$ underscreened local Kondo doublet $| \text{UKD}, m_{z}\rangle$ at site $i_{0}$ on orbital A and with the $T_{z}$ quantum number $m_{z}=\pm \sfrac12$, as indicated by the saturated correlation $\langle \hs \cdot \hS \rangle = -1$ (Fig.\ \ref{fig:under-pd}, left).

Furthermore, there are two excited $\Delta N=0$ subgap states, which become degenerate in the $J\to \infty$ limit (dark green in Fig.\ \ref{fig:under-pd}, upper right).
As a function of $J$, the first excited state evolves from the $J=0$ ground state, i.e., the product of the spin-singlet Dirac sea with the impurity-spin triplet. 
This state remains a total-spin triplet for strong $J$. 
Impurity-spin screening is only possible via processes involving particle-hole excitations across the gap. 

Tracing the second excited state, we find that this originates, at $J=0$, from the Dirac sea with an additional particle-hole excitation $c^{\dagger}_{\rm{cb}, \sigma} c_{\rm{vb}, \sigma'} | \mbox{sea} \rangle$ from the top of the valence band to the bottom of the conduction band with excitation energy $\Delta = 2$, in a product with the impurity-spin triplet. 
In the electronic sector, we thus have a singlet and a triplet state, which are degenerate at $J=0$. 
At weak $J > 0$, the electronic particle-hole triplet excitation and the impurity-spin triplet couple to total-spin singlet, triplet and quintuplet states, following the usual angular-momentum addition rules, i.e., $1 \otimes 1 = 0 \oplus 1 \oplus 2$. 
The total-spin singlet state has lowest energy.
Due to the additional electron in the conduction band, this second excited state supports a more efficient screening as compared to the first one, resulting in a fast decrease of its energy with $J$ (see the figure).

For $J>J_{\rm K}$, screening of the impurity spin is strong for both excited states. 
This is indicated by excitation energies $\Delta E_{1,2}(J) = E_{1,2}(J) - E_{0}(J)$ approaching a constant for $J\to \infty$, which implies $E_{1,2}(J) \simeq - J$ since $E_{0}(J) \simeq - J$ for $J\to \infty$.

In the $J\to\infty$ limit, the first two excited states, the total-spin triplet and the total-spin singlet, become degenerate and can be factorized into a tensor product of two doublets: $1 \oplus 0 = \sfrac12 \otimes \sfrac12$.
The first one is the underscreened local Kondo doublet, with energy $E_{\rm UKD} = m-J$. 
For $J \to \infty$, this state dynamically decouples from the rest of the system and is confined to the A orbital at site $i_{0}$ and the impurity spin.
For the rest of the system this ``underscreened Kondo hole'' acts as an effective spin-$\sfrac12$ impurity. 
This impurity binds the second factor, i.e., a degenerate doublet of ring states localized at, but ``outside'' the hole.

Analogous to the discussion in Sec.\ \ref{sec:top} for $S=\sfrac12$ and the local Kondo singlet hole, the bound ring-state doublet found for $S=1$ and the underscreened Kondo doublet, is a remnant of the chiral edge mode of the QWZ model in the k-space topologically nontrivial phase.
It may serve as a local signature of the (k-space) topological bulk phase, see Refs.\ \cite{SLLS11,SRZB15,DFVR20,MFQ+24}.

\begin{figure}[t] 
\centering
\includegraphics[width = 0.85\linewidth]{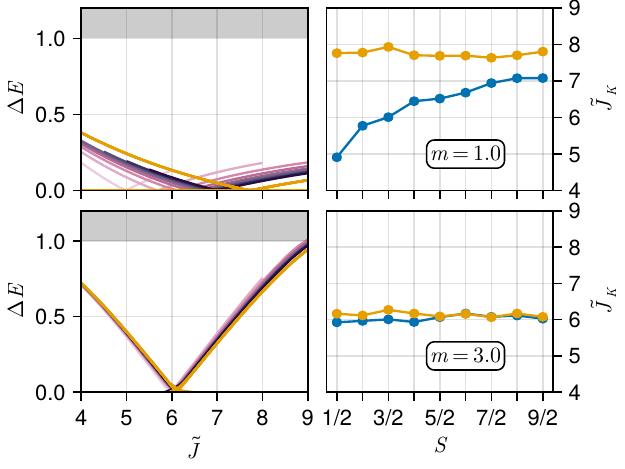}
\caption{
{\em Left:}
Flow of the difference $\Delta E$ between the first excited-state and the ground-state energy with $\tilde{J} \equiv J (S+1)$ for $S=\frac12 , 1, \frac32, ..., 4$ ({\em violet lines, from light to dark}).
{\em Yellow line}: the same but for a classical spin of length $\frac12$, but as a function of $\tilde{J} \equiv J S$.
{\em Right:}
Scaled critical coupling $\tilde{J_{\rm K}} \equiv J_{\rm K} (S+1) $ for $S=\frac12 , 1, \frac32, ..., 4$ ({\em blue circles}).
{\em Yellow circles}: Scaled critical coupling $\tilde{J_{\rm K}} \equiv J_{\rm K} S$ for a classical spin.
{\em Top}: $m=1$, {\em bottom}: $m=3$.
Results for a classical spin are obtained with $B \to \infty$.
}
\label{fig:under-flow}
\end{figure}

\subsection{Higher spin quantum numbers $S$}
\label{sec:highs}

For higher impurity-spin quantum numbers $S$ we expect that the deviations from the classical-spin results become progressively smaller and that for $S \to \infty$ one essentially recovers the results of the classical-spin system.
In fact, as demonstrated with Fig.\ \ref{fig:under-flow} (left panels), the difference between the energies of the lowest excited state and the ground state, $\Delta E (\tilde{J}) \equiv E_{1}(\tilde{J}) - E_{0}(\tilde{J})$ depending on the scaled coupling $\tilde{J} = J \,(S+1)$, appears to converge uniformly to the corresponding curve $\Delta E_{\rm class}(\tilde{J})$ with scaled coupling $\tilde{J} = S \, J$ for a classical spin.
We note that the convergence is much slower for the $k$-space topologically nontrivial case ($m=1$, top panel) as compared to the topologically trivial case ($m=3$, bottom panel).

The classical-spin results have been obtained by computations for a quantum impurity spin $S$, but with an additional local magnetic field $\ff B$ coupling to the impurity spin only and with field strength $B\to \infty$.
In practice $B=10^{4}$ is fully sufficient for convergence.
In the $B\to \infty$ limit, the remaining $S$ dependence is negligible, when rescaling the coupling as $J \to JS$ (note the rescaled axis in Fig.\ \ref{fig:under-flow}).
Actually this is enforced since, for $B\to \infty$, the interaction term $H_{J} = J \hs \cdot \hS \to J \hs \cdot \ff S = J S \hs \cdot \ff n$, where $\ff S \equiv S \ff n$ with $|\ff n|=1$, and thus the Hamiltonian depends on the product $JS$ only.
The scaling $J(S+1)$ for the quantum-spin results is motivated by \refeq{jk}, i.e., by the atomic limit.

From the zeros of $\Delta E (\tilde{J})$ for the various $S$, we can extract the $S$-dependent critical coupling $J_{\rm K}$.
Rescaled critical couplings $J_{\rm K} (S+1)$ and $J_{\rm K} S$ are shown in the right panels of Fig.\ \ref{fig:under-flow} as functions of $S$.
In the classical-spin case, one actually expects $J_{\rm K}S = \mbox{const.}$, as $H$ depends on the product $JS$ only. 
The remaining small deviations from a constant are numerical artifacts mainly resulting from the determination of the critical coupling from the data shown in the left panels of Fig.\ \ref{fig:under-flow}.

As a function of $S$, the (scaled) critical coupling $\tilde{J_{\rm K}} = J_{\rm K} (S+1)$ converges to the (scaled) critical coupling $\tilde{J}_{\rm K} = J_{\rm K} S$ for the classical-spin impurity, as demonstrated with the right panels in Fig.\ \ref{fig:under-flow}. 
Similar to the gap $\Delta E$ (left panels), we note that this convergence is much slower for the $k$-space topologically nontrivial ($m=1$) as compared to the trivial case ($m=3$).

\subsection{Insights from scattering theory}
\label{sec:scatt}

For a classical impurity spin, the Hamiltonian represents a system of {\em independent} electrons. 
The ``interaction term'' $H_{J} = J \hs \cdot \ff S$ is a simple local potential that induces electron scattering. 
The electronic structure and the critical interaction can thus be obtained by standard scattering theory, and the results should recover the numerical results obtained via the many-body approach in the classical-spin limit $B \to \infty$.
In addition, the scattering-theory approach is interesting, since, with suitable approximations, the same formalism can be employed for the quantum-spin problem as well. 

We start by considering a classical spin $\ff S = S \ff n$ with $|\ff n| = 1$ and reformulate the Hamiltonian $H=H_{0}+H_{J}$ in k-space as
\ba
H
&=& 
\sum_{\ff k \in \rm{BZ}} \sum_{r=\pm} \sum_{\sigma=\uparrow,\downarrow} \epsilon_{r}(\ff k) d^{\dagger}_{\ff k r \sigma} d_{\ff k r \sigma} 
\nonumber \\
&+&
J \sum_{\ff k\ff k'\sigma\sigma'} F(\ff k, \ff k')  d^{\dagger}_{\ff kr\sigma} \left(\frac{1}{2} \ff \sigma \cdot \ff S\right)_{\sigma\sigma'}  d_{\ff k'r'\sigma'}
\: .
\labeq{sc1}
\ea
Here, $d^{\dagger}_{\ff kr \sigma} = \sum_{\alpha=A,B} U_{\alpha r} (\ff k) c^{\dagger}_{\ff k \alpha \sigma}$, where $\ff U(\ff k)$ is the $\ff k$-dependent but $\sigma$-independent $2\times 2$ unitary matrix diagonalizing $H_{0}$, i.e., $\ff \varepsilon(\ff k)$.
Furthermore, we have defined $F(\ff k, \ff k') = U^{\dagger}_{Ar}(\ff k) e^{- i (\ff k - \ff k') \ff R_{i_{0}}} U_{r'A}(\ff k') / L$. 
This term contains the Fourier factor resulting from the transformation of $H_{J}$ to reciprocal space. 
Choosing $\ff R_{i_{0}} =0$, the latter reduces to $1/L$.

The poles of the Green's function matrix $\ff G(\omega) = \ff G^{(0)}(\omega) + \ff G^{(0)}(\omega) \ff T(\omega) \ff G^{(0)}(\omega)$ lying within the gap can be read off from the poles of the scattering T matrix. 
The latter is obtained from the free Green's function $\ff G^{(0)}(\omega)$ and the scattering potential $\ff V$ as 
\be
  \ff T(\omega) = \frac{\ff 1}{\ff 1 - \ff V \ff G^{(0)}(\omega)} \ff V
  \: .
\labeq{sc2}
\ee
The transition at $J=J_{\rm K}$ is characterized by a pole of $\ff T(\omega)$ at $\omega=0$ crossing $\mu=0$:
\be
  \det (\ff 1 - \ff V \ff G^{(0)}(\omega=0) ) \stackrel{!}{=} 0
  \: .
\labeq{cond}
\ee
Here, 
\ba
&&
G^{(0)}_{\ff kr \sigma, \ff k' r' \sigma'}(\omega) 
= 
\frac{\delta_{\ff k, \ff k'} \delta_{rr'} \delta_{\sigma\sigma'}}{\omega - \epsilon_{r}(\ff k)}
\:, 
\nonumber \\
&&
V_{\ff kr \sigma, \ff k' r' \sigma'}
= 
\frac{J}{L} 
U^{\dagger}_{Ar}(\ff k) 
\left(\frac{1}{2} \ff \sigma \cdot \ff S\right)_{\sigma\sigma'} 
U_{r'A}(\ff k') 
\labeq{sc3}
\: .
\ea

The condition \refeq{cond} can be evaluated conveniently using the matrix determinant lemma to get the determinant of a rank-one update of the unit matrix, 
\be
\det(\ff 1 + \ff u \ff v^{\dagger})  = 1 + \ff v^{\dagger} \ff u
\: ,
\labeq{lemma}
\ee
where $\ff u, \ff v$ are column vectors.
We have
\ba
&&
-(\ff V \ff G^{(0)})_{\ff kr \sigma, \ff k' r' \sigma'}(\omega=0) 
\nonumber \\
&&
= 
\frac{J}{L} 
U^{\dagger}_{Ar}(\ff k) 
\left(\frac{1}{2} \ff \sigma \cdot \ff S\right)_{\sigma\sigma'} 
U_{r'A}(\ff k') 
\frac{1}{\epsilon_{r'}(\ff k')}
\: .
\labeq{vg0}
\ea
Without loss of generality, we choose $\ff S$ to point in $z$ direction, $\ff S = S \ff e_{z}$, such that 
\be
\left(\frac{1}{2} \ff \sigma \cdot \ff S\right)_{\sigma\sigma'} = \frac12 S \delta_{\sigma\sigma'} z_{\sigma}
\: ,
\labeq{sc4}
\ee
with the sign $z_{\uparrow}=+1$ and $z_{\downarrow}=-1$.
With this choice the expression in \refeq{vg0} has the form of an outer product $\ff u \ff v^{\dagger}$, if we treat both spin channels $\sigma=\uparrow, \downarrow$ separately, i.e., if $\ff u$ has components $u_{\ff k r}$, and $\ff v$ analogously, and if $\sigma$ is fixed.
Defining
\ba
u_{\ff k r} =
\frac12 \frac{JS}{L} 
U^{\dagger}_{Ar}(\ff k) 
z_{\sigma}
\, ,  \;
v_{\ff k' r'} =
U^{\ast}_{r'A}(\ff k') 
\frac{1}{\epsilon_{r'}(\ff k')}
\: , 
\ea
we have $-(\ff V \ff G^{(0)})(\omega=0) = \ff u \ff v^{\dagger}$, and hence, 
using \refeq{lemma}, the condition \refeq{cond} reads:
\be
0 \stackrel{!}{=} 1 + \ff v^{\dagger} \ff u
=
1 + 
\sum_{\ff k r}
U_{rA}(\ff k) 
\frac{1}{\epsilon_{r}(\ff k)}
\frac12 \frac{JS}{L} 
U^{\dagger}_{Ar}(\ff k) 
z_{\sigma}
\: .
\ee
For antiferromagnetic coupling $J>0$, the dominant channel is given by $z_{\sigma}=-1$, or, equivalently, for $\sigma=\downarrow$, i.e., if the electron spin and the impurity spin align antiparallel. 
Using this and solving for $J$ yields the critical interaction:
\be
J_{\rm K}
=
\frac{2}{S} 
\left(
\frac{1}{L} \sum_{\ff k} \sum_{r} \frac{| U_{rA}(\ff k) |^{2} }{\epsilon_{r}(\ff k)}
\right)^{-1}
\: .
\labeq{jkcl}
\ee

\begin{figure}[t] 
\centering
\includegraphics[width = 0.75\linewidth]{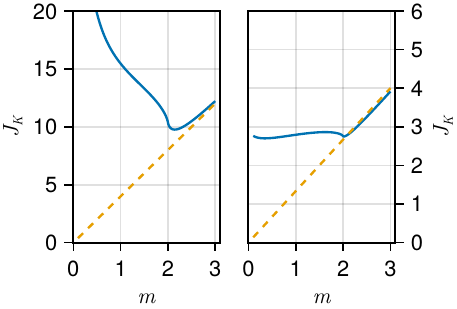}
\caption{
{\em Left:}
$J_{\rm K}(m)$ as obtained by scattering theory, Eq.\ (\ref{eq:jkcl}), for a classical spin (length $S=1/2$, prefactor $2/S= 4$).
{\em Right:}
$J_{\rm K}(m)$ from projected scattering theory, Eq.\ (\ref{eq:jkqu}), for a quantum spin $S=\sfrac12$, (prefactor $2/(S+1)= \sfrac43$).
The respective prefactor determines the slope of the linear trend at large $m \gtrsim 2$ (dashed lines).
}
\label{fig:scjk}
\end{figure}

The integral over the Brillouin zone in \refeq{jkcl} is readily evaluated numerically. 
The resulting $m$ dependence of the critical coupling $J_{\rm K}$ is shown in the right panel of Fig.\ \ref{fig:scjk} for a spin of length $S=\sfrac12$.
Comparing with the result obtained by the full many-body approach but for $B \to \infty$, see Fig.\ \ref{fig:pdclass} (left), there is perfect agreement, as expected, since the scattering theory is exact. 

\begin{figure}[b] 
\centering
\includegraphics[width = 0.5\linewidth]{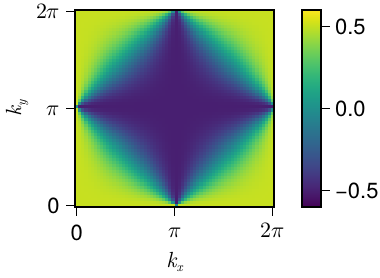}
\caption{
Integrand $I(\ff k)$ in Eq.\ (\ref{eq:jkcl}) for $m=0$.
We have $I(\ff k) = - I(\ff k + \ff Q)$ with $\ff Q = (\pi, \pi)$.
}
\label{fig:scint}
\end{figure}

For $m\to 0$, the critical coupling diverges as $J_{\rm K} \sim - 1 / (m \ln m)$, as determined numerically. 
The divergence at $m=0$ is easily seen to result from the antisymmetry $\ff \epsilon(\ff k) = - \ff \epsilon(\ff k + \ff \pi)$ of the Bloch Hamiltonian, \refeq{bloch}, under a shift by $\ff \pi = (\pi, \pi)$. 
Using the notation $\ff \epsilon(\ff k) = \ff d(\ff k) \cdot \ff \tau$ with $\ff d(\ff k) = (t \sin k_x , t \sin k_y , m+ t \cos k_x + t \cos k_y)$, 
we have $\ff \epsilon(\ff k)^{-1} = \ff d(\ff k) \cdot \ff \tau / d(\ff k)^{2}$. 
For the integrand $I(\ff k)$ of the $\ff k$ integration over the BZ in \refeq{jkcl} this implies the same antisymmetry (for $m=0$),
\ba
I(\ff k) & \equiv &
\sum_{r} \frac{| U_{rA}(\ff k) |^{2} }{\epsilon_{r}(\ff k)}
=
\langle \ff k A | \frac{1}{\ff \epsilon(\ff k) } | \ff k A \rangle 
\nonumber \\
& = &
\frac{\langle \ff k A | \ff d(\ff k) \cdot \ff \tau | \ff k A \rangle }{ d(\ff k)^{2} } 
=
\frac{d_{z}(\ff k)}{ d(\ff k)^{2} } 
= - I(\ff k + \ff \pi) 
\: , 
\nonumber \\
\labeq{ik}
\ea
see also Fig.\ \ref{fig:scint}. 
As a consequence, the integral over the BZ vanishes, i.e., $J_{\rm K}(m=0)$ diverges.
For all $m \ne 0$, the antisymmetry is lost and $J_{\rm K}$ remains finite.

As discussed in Sec.\ \ref{sec:cl}, the critical coupling approaches $J_{\rm K}=4m$ for a classical spin of length $S=\sfrac12$ if $m$ is large. 
This result is recovered with \refeq{jkcl}, numerically (Fig.\ \ref{fig:scjk}), and analytically within the atomic limit:
Setting $t=0$, we have $\ff d(\ff k) = m \ff e_{z}$ and thus, via \refeq{ik}, $I(\ff k) = 1/m$, which yields $J_{\rm K} = 2m / S = 4m$ for $S=\sfrac12$. 
This is indicated by the dashed line in Fig.\ \ref{fig:scjk} (left).

For a quantum spin $S$, the scattering-theory approach can be set up in essentially the same way. 
Equations (\ref{eq:sc1}) -- (\ref{eq:vg0}) remain valid with the replacement of $\ff S$ by $\hS$.
However, Eq.\ (\ref{eq:sc4}) no longer applies.
Instead, we exploit the fact that $\ff \sigma$ and $\hS$ act on different (electron and impurity-spin) sectors and use the identity
\be
\ff \sigma \cdot \hS 
=
\left( \frac12 \ff \sigma + \hS \right)^{2} - \frac{3}{4} - S (S+1)
\: .
\labeq{sc4n}
\ee
We then adopt the approximation $(\ff \sigma/2 + \hS)^{2} = T ( T + 1)$ with $T = S-\sfrac12$, thereby assuming that the electron spin 
and the impurity spin couple antiferromagnetically.
While this is physically plausible, it neglects quantum fluctuations. 
Virtual processes breaking up and restoring the antiferromagnetic coupling may decrease the ground-state energy to some extent.
Within the approximation and with $T ( T + 1) = (S-\sfrac12)(S+\sfrac12) = S^{2} - \sfrac14$, we can simply replace
\be
\left(\frac{1}{2} \ff \sigma \cdot \hS \right)_{\sigma\sigma'} 
\mapsto
- \frac{S +1}{2} \delta_{\sigma\sigma'}
\labeq{approx}
\ee
in \refeq{vg0}.
The rest of the calculation proceeds analogously to the classical-spin case and results in
\be
J_{\rm K}
=
\frac{2}{S+1} 
\left(
\frac{1}{L} \sum_{\ff k} \sum_{r} \frac{| U_{rA}(\ff k) |^{2} }{\epsilon_{r}(\ff k)}
\right)^{-1}
\: .
\labeq{jkquaf}
\ee
This differs from the result for the classical-spin case, \refeq{jkcl}, by the prefactor $1/(S+1)$ only. 
For $S \to \infty$, when the classical-spin approximation is expected as justified, this is irrelevant. 
The change is important, however, when comparing the approximate scattering theory, \refeq{jkquaf}, with the full many-body theory: 
The latter predicts the rescaled critical coupling $J_{\rm K}(S) \cdot (S+1)$ as basically independent of $S$ for $m=3$
and, for $m=1$, to approach the classical-spin limit significantly faster, only if the correct scaling factor $(S+1)$ is taken into account, as is demonstrated in Fig.\ \ref{fig:under-flow} (right).
For smaller $S$, the trend of $J_{\rm K}$ with $S$ is seen to increasingly deviate from the predicted $\propto 1/(S+1)$ trend.

Evaluating \refeq{jkquaf} in the atomic limit is instructive again:
For $t=0$, we find $J_{\rm K} = 2m / (S+1)$ which nicely recovers the many-body-theory results in the regime $m \gtrsim 2$ for a quantum spin $S=1$ ($J_{\rm K} = m$) and even for a quantum spin $S=\sfrac12$ ($J_{\rm K} = 4m/3$), see Figs.\ \ref{fig:under-pd} and \ref{fig:pd1}, respectively.

In the $m \lesssim 2$ regime, however, the limitations of the approximation \refeq{approx} become apparent.
Most importantly, \refeq{jkquaf} predicts a diverging $J_{\rm K}(m)$ for $m\to 0$, whereas $J_{\rm K}$ approaches a constant value for spin quantum number $S=\sfrac12$ (and, less obviously, for $S=1$ as well).
Furthermore, it cannot reproduce the nearly constant trend of $J_{\rm K}$ with $m$ in the $m \lesssim 2$ regime.

For the extreme case of a quantum spin $S=\sfrac12$, an improved variant of the scattering-theory approach is available.
The central approximation is to project the problem onto the conduction-band ($r=+$) subspace and to determine the critical interaction $J_{\rm K}$ from the resulting projected theory. 
Operationally, we again start from Eqs.\ (\ref{eq:sc1}) -- (\ref{eq:sc3}), with $\ff S$ replaced by $\hS$, but restrict the matrix elements of $\ff G^{(0)}$ and $\ff V$ in \refeq{sc3} to $r=r'=+$.
Within this approximation the completely filled valence band is treated as inert.

In terms of many-body states, this approximation means that the ground state in the $\Delta N=0$ sector is given by a completely filled valence band and an empty conduction band without virtual fluctuations across the gap, i.e., virtual screening of the impurity spin is neglected.
As compared to $S=1$, virtual screening is in fact strongly reduced in the $S=\sfrac12$ case: 
In the relevant regime $m<2$ and for $m\to 0$ in particular, the local spin correlation $\langle \hs \cdot \hS \rangle$, relative to its saturation value, has a considerably smaller modulus in the $S=\sfrac12$ case, see Figs.\ \ref{fig:pd1} and \ref{fig:under-pd}.
The main underlying reason for the suppressed virtual screening is that, as compared to the case of $S=1$, the exchange-energy gain is considerably smaller for the lower spin quantum number $S=\sfrac12$.
Since the impurity spin is ``shorter'', it generates a weaker effective local magnetic field, so that the screening electron gains less exchange energy upon aligning with it.

Neglecting virtual screening in the $\Delta N=0$ sector implies that screening of the $S=\sfrac12$ impurity spin can only occur through an electron-addition excitation. 
This is precisely the screening mechanism captured by the scattering theory based on the single-particle Green's function. 
Since the approximation eliminates virtual screening, one expects the resulting critical coupling $J_{\rm K}$ to be reduced relative to the exact many-body result, thereby partially compensating for the neglected screening mechanism.

As for the many-body ground state in the $\Delta N=1$ sector, the projection is essentially exact, since the additional conduction-band electron forms a singlet with the impurity spin $S=\sfrac12$. 
Consequently, virtual fluctuations across the gap are ineffective and do not produce any further exchange-energy gain. 
In particular, once virtual interband processes have been neglected, the replacement in Eq.\ (\ref{eq:approx}) becomes exact for the $\Delta N=1$ ground state.

The projected theory therefore yields the following expression for the critical interaction:
\be
J_{\rm K}
=
\frac{4}{3} 
\left(
\frac{1}{L} \sum_{\ff k} \frac{| U_{+,A}(\ff k) |^{2} }{\epsilon_{+}(\ff k)}
\right)^{-1}
\: .
\labeq{jkqu}
\ee
This differs from \refeq{jkquaf} in two respects: 
first, the contribution of the $r=-$ band is omitted and, second, the result is specific to the case $S=\sfrac12$.

As a simple check, we specialize to the atomic limit $t=0$, where $|U_{rA}(\ff k)|^{2} = \delta_{r,+}$ and $\epsilon_{+}(\ff k) = m$. 
This implies $J_{\rm K} = 4m/3$.
For $S=\sfrac12$ and $m \gtrsim 2$, this exactly recovers the result $J_{\rm K} \simeq \frac43 m$.
Note that this is the same as obtained with \refeq{jkquaf}. 

The critical coupling $J_{\rm K}(m)$ as derived by numerical evaluation of \refeq{jkqu} is shown in the right panel of Fig.\ \ref{fig:scjk}.
The dashed line indicates the $J_{\rm K} \simeq \frac43 m$ trend. 
In the $m<2$ regime, $J_{\rm K}(m)$ is roughly constant and thus in nice qualitative agreement with the full $S=\sfrac12$ phase diagram shown in Fig.\ \ref{fig:pd1}. 
For $m\to 0$, the critical coupling remains finite. 
However, it does show a slight upturn in this limit that is also seen in the many-body approach (Fig.\ \ref{fig:pd1}).
Quantitatively, differences are to be expected, since many-body effects, such as virtual screening in the $\Delta N=0$ sector, are neglected. 
Consequently, \refeq{jkqu} must underestimate the correct value of $J_{\rm K}$ for $m<2$. 
In fact, comparing with Fig.\ \ref{fig:pd1} shows that it is predicted to be about 20\% too small.

\subsection{Overscreened case}
\label{sec:over}

We replace the interaction term $H_{J}$ in \refeq{ham} by $H_{J} = J_{\rm A} \hS \cdot \hs_{i_{0}A} + J_{\rm B} \hS \cdot \hs_{i_{0}B}$, i.e., we couple the impurity spin to the local electron spins on both orbitals at $i_{0}$. 
In this overscreened, two-channel \cite{NB80,SHPM15} case, for strong $J$, there are two competing ways to locally screen the impurity spin, namely singlet formation with either $\hs_{i_{0}A}$ or $\hs_{i_{0}B}$. 
As discussed in the following, this leads to an intricate phase diagram and, in particular, to a quantum-classical crossover governed by spontaneous particle-hole symmetry breaking.
For equal coupling strengths $J \equiv J_{\rm A}=J_{\rm B}$, the Hamiltonian $H_{0} + H_{J}$ is invariant under the particle-hole (PH) transformation that is unitarily represented on the fermion Fock space by $\ca P$ acting as $c_{i \alpha \sigma} \mapsto \ca P c_{i \alpha \sigma} \ca P^{\dagger} \equiv c^{\dagger}_{i \overline{\alpha} \sigma}$ where $\overline{\alpha} = \mbox{B}$ if $\alpha = \mbox{A}$ and vice versa, and as $\hS \mapsto \ca P \hS \ca P^{\dagger} \equiv (- \hat S_{x}, \hat S_{y}, -\hat S_{z})$.

\begin{figure}[t] 
\centering
\includegraphics[width = 0.69\linewidth]{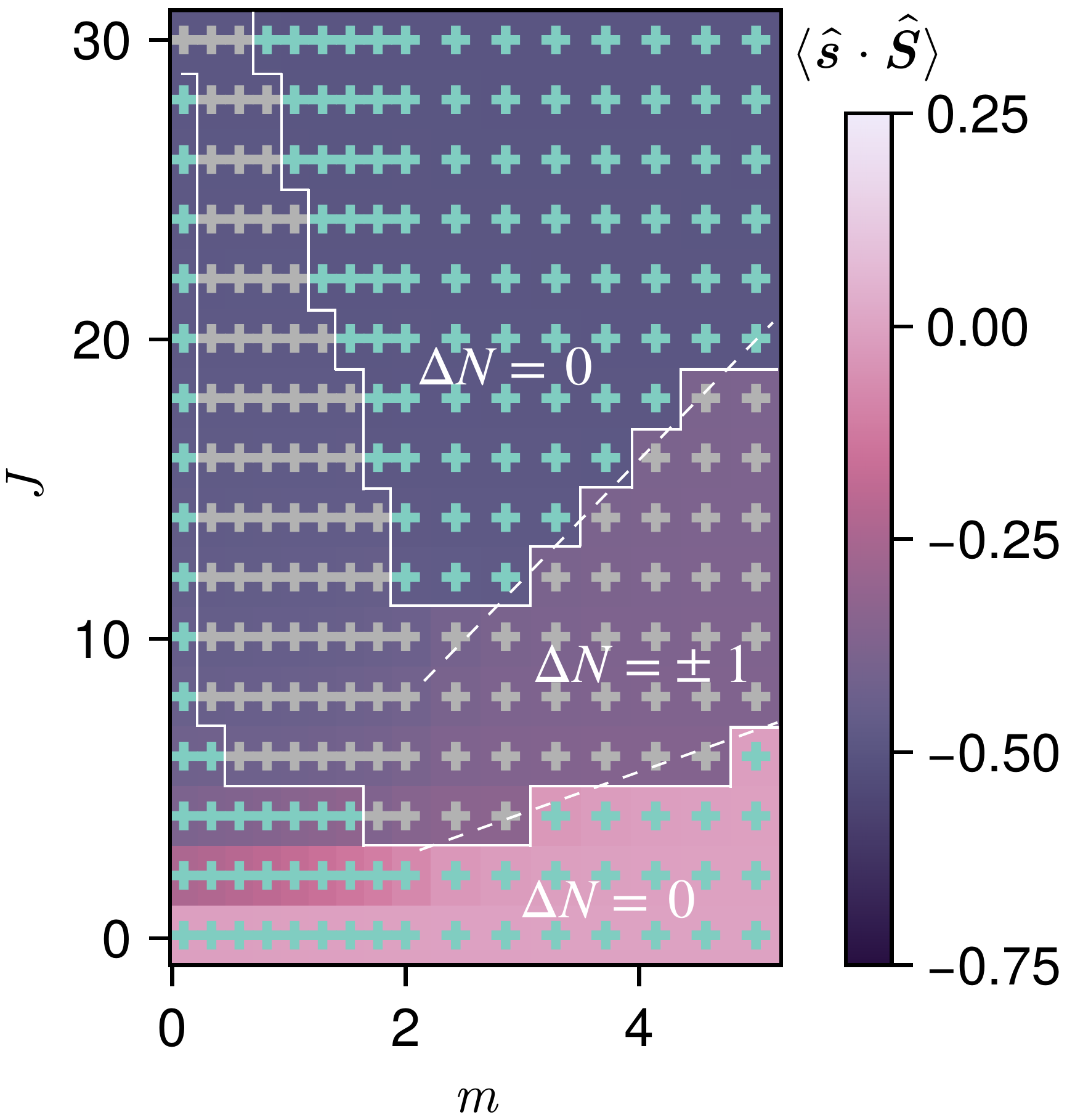}
\caption{
$m$-$J$ phase diagram as obtained numerically for the overscreened quantum-spin impurity in the symmetric case $J=J_{\rm A} = J_{\rm B}$. 
{\em White lines:} discontinuous phase transitions.
{\em Color code:} local spin correlations $\langle \hs \cdot \hS \rangle \equiv \langle \hs_{i_{0}A} \cdot \hS \rangle = \langle \hs_{i_{0}B} \cdot \hS \rangle$ (for $\Delta N = \pm 1$ averaged over the two sectors).
{\em Dashed white lines:} see text.
{\em Light blue symbol color:} $\Delta N = 0$, {\em gray:} $\Delta N = \pm 1$. 
{\em Symbol form:} ground-state degeneracy. Two singlets: {\em cross}, singlet: {\em circle}. 
Calculations for $N = 102 + \Delta N$.
}
\label{fig:mover}
\end{figure}

We first discuss the quantum-spin case.
For $J=0$ we trivially have $\cb=-1$. 
In the $J\to \infty$ limit, the occupations $\langle n_{i_{0}A\sigma} \rangle, \langle n_{i_{0}B\sigma} \rangle \to 0.5$, and the ground state of $H_{J}$ and thus of $H_{0} + H_{J}$ is a total-spin doublet with spin correlations $\langle \hs_{i_{0}A} \cdot \hS \rangle = \langle \hs_{i_{0}B} \cdot \hS \rangle \to -1/2$.
Coupling this doublet perturbatively to the local magnetic field $B\searrow 0$, again yields $\cb = -1$.
However, the numerically obtained $m$-$J$ phase diagram, see Fig.\ \ref{fig:mover}, shows the presence of {\em two} critical couplings $J_{\rm c1}(m) < J_{\rm c2}(m)$ (white lines) instead of none.
For $J_{\rm c1}(m) < J < J_{\rm c2}(m)$, and in the entire $m$ range, we find an intermediate phase with two degenerate ground states in the sectors with $\Delta N = \pm 1$.

For large $m$ and thus for strong orbital polarization, its extension increases linearly with increasing $m$.
This trend can be easily understood within an effective atomic-limit ($t=0$) model: 
Comparing the ground-state energy for weak $J$ in the $\Delta N=0$ sector, $E_{\rm 2e} =-2m$ (fully occupied B orbital), with the ground-state energy $E_{\rm 1e} = - \sfrac34 J - m$ in the $\Delta N = -1$ sector (local singlet with the singly occupied B orbital), yields $J_{\rm c1} = \sfrac43 \, m$. 
We see that, for $J \ll m$, orbital polarization wins over local Kondo-singlet formation. 

The ground-state energy in the $\Delta N = +1$ sector (local singlet with the singly occupied A orbital, doubly occupied B orbital) is the same as in the $\Delta N = -1$ sector, $E_{\rm 3e} = - \sfrac34 J - 2m + m = E_{\rm 1e}$. 
To find $J_{\rm c2}$, we again compare with the ground-state energy in the $\Delta N=0$ sector. 
However, both orbitals are now singly occupied since developing local antiferromagnetic Kondo correlations becomes favorable as compared to orbital polarization in the strong-$J$ case.
This leads to an effective spin model $H_{\rm eff} = - J \hs_{i_{0}A} \cdot \hS - J \hs_{i_{0}B} \cdot \hS$ with antiferromagnetic ground-state spin correlations $\langle \hs_{i_{0} A} \cdot \hS \rangle = \langle \hs_{i_{0} B} \cdot \hS \rangle = - \frac12 > - \frac34$. 
Hence, the $\Delta N=0$ ground-state energy is now given by $E_{\rm 2e} = - \langle \hs_{i_{0} A} \cdot \hS \rangle J - \langle \hs_{i_{0} B} \cdot \hS \rangle J = - 2 J/2 = - J$. 
Comparing with $E_{\rm 3e}$ this yields $J_{\rm c2} = 4 m$. 
In Fig.\ \ref{fig:mover}, the critical couplings $J_{\rm c1} = \sfrac43 \, m$ and $J_{\rm c2} = 4m$ are indicated as dashed white lines. 
We see that these almost perfectly describe the numerically obtained boundaries of the intermediate phase in the strong-$m$ limit.

For smaller mass parameters $m$, the critical couplings $J_{\rm c1}(m)$ and $J_{\rm c2}(m)$ become strongly $m$ dependent, and in the limit $m \to 0$ they even appear to diverge. 
This trend cannot be easily deduced from a simple effective model.
A reliable numerical study of this limit is challenging, too, since the insulating gap shrinks to zero for $m \to 0$ such that finite-size effects cannot be controlled efficiently. 

Fig.\ \ref{fig:mover} shows that the intermediate $\Delta N=\pm 1$ phase for moderate $m$ smoothly connects to the one for large $m$.
However, one finds qualitatively different correlations:
Consider the spin correlation $\langle \hs \cdot \hS \rangle \equiv \langle \hs_{i_{0}A} \cdot \hS \rangle = \langle \hs_{i_{0}B} \cdot \hS \rangle$ that is obtained by arithmetically averaging over the two $\Delta N=\pm 1$ sectors. 
For $J=12$, we numerically find
$\langle \hs \cdot \hS \rangle \approx -0.46$ at $m=1$, while at $m=5$ we get 
$\langle \hs \cdot \hS \rangle \approx -0.37 \approx -\sfrac38$.
The strong-$m$ correlation close to $\langle \hs \cdot \hS \rangle = - \sfrac38$ results from local Kondo-singlet formation either with $\hs_{i_{0}A}$ ($\Delta N = +1$) or with $\hs_{i_{0}B}$ ($\Delta N = -1$):
In the $\Delta N=+1$ ground state, the impurity spin $\hS$ develops strong antiferromagnetic correlations with the 
additional electron mainly localized on the A orbital at $i_{0}$, while the removal of an electron, mainly from the B orbital at $i_{0}$ ($\Delta N=-1$), leads to strong antiferromagnetic correlations of $\hS$ with the remaining electron in B. 

For smaller mass parameters, e.g., $m=1$, however, the situation is different since the electrons are more delocalized. 
In fact, for the sector $\Delta N=+1$, it turns out that it is energetically favorable to decrease the electron number {\em locally} at the site $i_{0}$ by one (and, vice versa, for $\Delta N=-1$ to increase it by one), such that in both cases there are two electrons at $i_{0}$, one localized on the A and one on the B orbital and forming local spins $\sfrac12$. 
This results in $\langle \hs \cdot \hS \rangle \approx - 0.5 < - \sfrac38$ and thus in a stronger negative contribution to the ground-state energy.
It also implies that, locally, the impurity spin, together with well-developed spins $\sfrac12$ on the A and the B orbital  forms a total spin $\sfrac12$.
The total spin $\sfrac12$ at $i_{0}$ is screened by the remaining (odd number of) electrons in the rest of the system, excluding site $i_{0}$. 
As a result, a total-spin singlet state is formed, one in the sector $\Delta N=+1$ and a degenerate one in the sector $\Delta N =-1$.
We conclude that the intermediate $\Delta N = \pm 1$ phase for both large and small $m$, has B-space Chern number $\cb=0$, since a weak field $\ff B$ cannot couple to a singlet, neither the one localized at $i_{0}$ formed with a single orbital nor the more delocalized one formed by the local total spin $\sfrac12$ at $i_{0}$ with the conduction electrons in the rest of the system.

\subsection{Quantum-classical deformation and spontaneous symmetry breaking}
\label{sec:ssb}

\begin{figure}[t] 
\centering
\includegraphics[width = 0.65\linewidth]{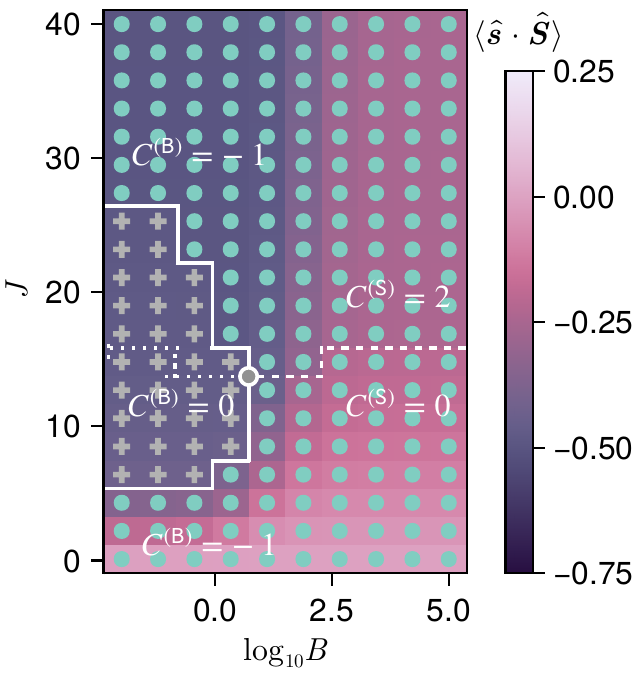}
\\
\includegraphics[width = 0.65\linewidth]{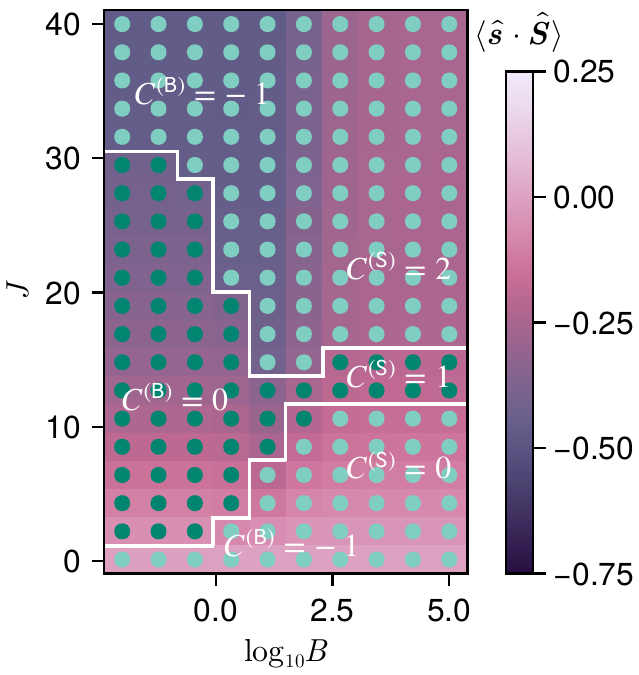}
\caption{
$B$-$J$ phase diagrams in the overscreened case for $m=1$ and $J=J_{\rm A} = J_{\rm B}$ {\em (top)} and for $J_{\rm A} = J$ and $J_{\rm B} = J + \Delta J$ with $\Delta J = 3$ {\em (bottom)}. 
{\em Color code:} local spin correlations $\langle \hs \cdot \hS \rangle \equiv \langle \hs_{i_{0}A} \cdot \hS \rangle = \langle \hs_{i_{0}B} \cdot \hS \rangle$ (for $\Delta N = \pm 1$ averaged over the two sectors).
{\em Light blue symbol color:} $\Delta N = 0$, 
{\em gray:} $\Delta N = \pm 1$. 
{\em green:} $\Delta N = - 1$. 
{\em Symbol form:} ground-state degeneracy $g$. $g=2$: {\em cross}, $g=1$: {\em circle}. 
{\em Full white lines:} spontaneous PH symmetry-breaking phase transition {\em (top)} or topological transition {\em (bottom panel)}.
{\em Dashed line:} topological transition between $\cs=0$ and $\cs=2$ {\em (top)}. 
{\em Dotted line:} transition in the metastable $\Delta N=0$ sector {\em (top)}. 
{\em White circle:} critical point. 
Calculations for $N = 102 + \Delta N$.
}
\label{fig:bover}
\end{figure}

The $B$-$J$ phase diagram for the quantum-spin-classical-spin crossover at fixed $m=1$ is shown in Fig.\ \ref{fig:bover}. 
We first discuss the particle-hole symmetric model with $J=J_{\rm A} = J_{\rm B}$, see the top panel of the figure.
In the classical-spin case, i.e., in the $B\to \infty$ limit, we trivially have $\cs = 0$ if $J=0$.
For $J \to \infty$, the classical ``field'' $J \ff S$ aligns $\hs_{i_{0}A}$ and $\hs_{i_{0}B}$ ferromagnetically. 
The equivalent effective problem is that of a quantum spin $S=1$ in an external magnetic field. 
This has an S-space Chern number $\cs = 2$ (see Ref.\ \cite{Ber84} for a concrete calculation).
We conclude that there must be a critical coupling $J_{\rm c}(m)$ in the classical limit, contrary to the quantum-spin case, where a gap closure is not {\em enforced} as $\cb=-1$ for both $J=0$ and $J=\infty$. 
However, as discussed above there is an intermediate $\Delta N=\pm 1$ phase with B-space Chern number $\cb=0$. 
Here, in the $B$-$J$ phase diagram, the respective critical couplings $J_{\rm c1}(B) < J < J_{\rm c2}(B)$ become $B$ dependent, see white lines in Fig.\ \ref{fig:bover} (top).

In the quantum-spin case, $B\to 0$, the two degenerate ground states in the sectors $\Delta N = \pm 1$ are not invariant under the PH transformation but transform into each other. 
Since the Hamiltonian is invariant, this implies that PH symmetry is broken {\em spontaneously}. 
On the other hand, in the classical-spin case for $B\to \infty$, the ground state is unique and PH symmetric. 
This is plausible, since SU(2) spin-rotation symmetry is broken explicitly for a classical impurity spin $\ff S$, and {\em both} local electron spins, $\hs_{i_{0}A}$ and $\hs_{i_{0}B}$, can fully align antiferromagnetically with $\ff S$ without any competition. 
In the quantum-spin case this is impossible.
To gain as much exchange energy as possible, PH-symmetry breaking turns out to be favorable in a certain $J$ regime, since this leads to almost singly occupied A or B orbitals, such that singlet formation of $\hS$ with either $\hs_{i_{0}A}$ or $\hs_{i_{0}B}$ becomes possible for large $m$, or, for small $m$, the formation of a local spin-doublet state with $\hS$ antiferromagnetically coupled to both $\hs_{i_{0}A}$ and $\hs_{i_{0}B}$.

For the quantum-classical crossover, this means that there must be a critical field $B_{\rm c}$ above which a unique and PH-symmetric ground state is restored. 
We find that the critical field is finite and marks a local phase transition. 
As is seen in Fig.\ \ref{fig:bover} (top), the intermediate symmetry-broken phase shrinks with increasing $B$, and the $B$-dependent critical couplings $J_{\rm c1}(B)$ and $J_{\rm c2}(B)$ continuously merge at a critical point $B=B_{\rm c}$, where $J_{c1}(B_{\rm c}) = J_{c2}(B_{\rm c})$, indicated by the white circle in the figure.
For $B>B_{\rm c}$, there is only a single critical line $J_{\rm c}(B)$ along which the gap is closed.
This line marks an almost $B$-independent transition, see dashed white line in Fig.\ \ref{fig:bover} (top).
When tracing its fate for $B \to \infty$, it turns out to end exactly at the transition point $J_{\rm c}(\infty)$ that separates the $\cs=0$ phase from the $\cs=2$ phase of the classical-spin case.
All in all, this demonstrates that the quantum-classical crossover is {\em not} continuous. 

It is conceivable, however, that a transition between phases, which are topologically characterized by S-space Chern numbers $\cs=0$ and $\cs=2$, could take place via an intermediate $\cs = 1$ phase.
In fact, this is the scenario that is realized in the asymmetric case $J_{\rm A} < J_{\rm B}$. 
As is demonstrated in Fig.\ \ref{fig:bover} (bottom), in the strong-$B$ regime, the single critical line $J_{\rm c}(B)$ splits up into two separate phase-transition lines $J_{\rm c1}(B) < J_{\rm c2}(B)$, which smoothly connect to the corresponding critical points $J_{\rm c1}$ and $J_{\rm c2}$ of the quantum-spin case.
For $J_{\rm A} < J_{\rm B}$, however, the intermediate phase with 
$J_{\rm c1} < J_{\rm A} < J_{\rm c2}$ and $J_{\rm B} = J_{\rm A} + \Delta J$ (here: $\Delta J = 3$) is no longer characterized as a spontaneously symmetry-broken phase, since $J_{\rm A} < J_{\rm B}$ explicitly breaks the PH symmetry of the Hamiltonian.
In the asymmetric case, the quantum-classical crossover is {\em continuous} again.

One may also constrain the system to remain in the $\Delta N=0$ ground state for $B < B_{\rm c}$. 
Tracing the (dashed) transition line with decreasing $B$ beyond $B_{\rm c}$, i.e., into the metastable $\Delta N=0$ regime for $B < B_{\rm c}$, see the dotted white line in Fig.\ \ref{fig:bover} (top), one finds a critical coupling $J_{\rm c} \approx 15$ at $B=0$, as defined by a gap closure.
This critical point $J_{\rm c}$ separates two metastable phases that can still be characterized by $\cs=0$ and $\cs = 2$ using a continuity argument.
However, the physical meaning of the S-space Chern number in the full quantum-spin limit is elusive. 
The transition at $J_{\rm c}$ is a nontopological level crossing between two metastable states, which continuously connect to the weak-$J$ and the strong-$J$ phases in the $\Delta N=0$ sector and which both carry the same B-space Chern number $\cb=-1$.
Kondo screening, however, is achieved in different ways: 
In the $\Delta N=0$ metastable ground state connected to the weak-$J$ regime, Kondo-singlet formation is mainly due to virtual fluctuations across the gap such that temporarily the almost doubly occupied B orbital can contribute to screening.
On the other hand, in the state connected to the strong-$J$ regime, $\hS$ develops strong antiferromagnetic correlations with both $\hs_{i_{0},A}$ and $\hs_{i_{0},B}$.

Let us return to the classical-spin case and discuss the role of PH symmetry for the topological phase diagram by analyzing the underlying electronic structure. 
For the PH-symmetric case $J_{\rm A} = J_{\rm B}$, the entire spectrum of the classical-spin and thus non-interacting Hamiltonian is symmetric with respect to $\mu=0$. 
In particular, this implies that, when varying $J$, an in-gap single-electron state crossing $\mu$ from above (below) must have a PH partner state with exactly the opposite energy crossing $\mu$ from below (above). 
Therefore, both states cross $\mu$ at the same critical interaction $J_{\rm c}$. 
Both states carry opposite spin-projection quantum numbers $\sigma=\uparrow, \downarrow$ and thus opposite single-particle S-space Chern numbers $\scs=\pm 1$, 
so that the total S-space Chern number must exactly change by $\Delta \cs = 2$.
If, on the other hand, $J_{\rm A} \ne J_{\rm B}$ and PH symmetry is broken, both states will generically cross $\mu=0$ at different interaction strengths $J_{\rm c1} \ne J_{\rm c2}$.
This necessitates the existence of an intermediate $\cs=1$ 
phase -- as is indeed found, see Fig.\ \ref{fig:bover} (bottom) in the classical-spin limit $B\to \infty$.

\begin{figure}[t] 
\centering
\includegraphics[width = 0.8\linewidth]{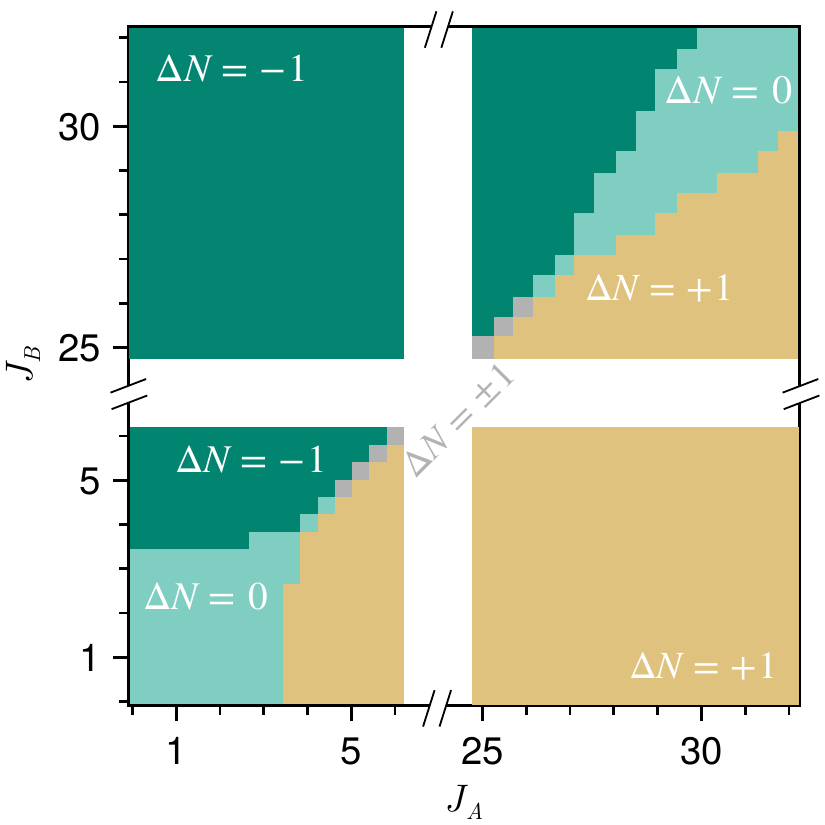}
\caption{
Quantum-impurity-spin $J_{A}$-$J_{B}$ phase diagram at $m=1$.
{\em Light green}: phases with nondegenerate singlet ground state in the $\Delta N = 0$ sector, $\cb=-1$. 
{\em Green/yellow}: ground states in the $\Delta N = -1$ / $\Delta N =+ 1$ sectors, respectively. $\cb=0$. 
{\em Gray}: degenerate ground states in the $\Delta N = \pm 1$ sectors, $\cb=0$. 
Calculations for $N = 102 + \Delta N$.
}
\label{fig:jajb}
\end{figure}

In the quantum-spin case, the situation is more complicated due to the PH-symmetry breaking. 
We find four different phases when varying $J_{\rm A}$ and $J_{\rm B}$ independently from zero to infinity, as demonstrated in Fig.\ \ref{fig:jajb}.
For an understanding of the structure of the phase diagram, it is again instructive to determine the B-space Chern number $\cb$. 
As discussed before, we have $\cb = -1$ in the PH-symmetric case $J_{\rm A} = J_{\rm B}$ in both limits, $(J_{\rm A}, J_{\rm B})=(0,0)$ and $(J_{\rm A}, J_{\rm B})=(\infty,\infty)$, and thus there is no phase transition enforced. 
Numerically, however, we have identified the intermediate $\cb=0$ phase, which is due to the emergent PH-symmetry breaking and which is characterized by two degenerate singlet ground states in different sectors $\Delta N = \pm 1$, see the gray phase on the diagonal $J_{\rm A}=J_{\rm B}$ in Fig.\ \ref{fig:jajb}, which separates the two weak- and the strong-coupling $\Delta N = 0$ phases with $\cb=-1$.

Considering the ``bottom-right'' path $(J_{\rm A}, J_{\rm B}) = (0,0) \mapsto (\infty, 0) \mapsto (\infty, \infty)$, rather than moving along the diagonal, immediately gives us two critical points, $(J_{1},0)$ and $(\infty, J_{2})$ along the $J_{\rm B}=0$ and along the $J_{\rm A}=\infty$ line.
These are enforced by different Chern numbers $\cb = -1$ at $(J_{\rm A}, J_{\rm B}) = (0,0)$ and $\cb = 0$ at $(\infty,0)$ (local Kondo singlet with $\hs_{i_{0},A}$) and $\cb = -1$ at $(J_{\rm A}, J_{\rm B}) = (\infty,\infty)$. 
Vice versa, there are two critical points $(0,J_{1})$ and $(J_{2},\infty)$ along the ``left-top'' path 
$(J_{\rm A}, J_{\rm B}) = (0,0) \mapsto (0,\infty) \mapsto (\infty, \infty)$. 

Note that the ground states at $(J_{\rm A}, J_{\rm B})$ and at $(J_{\rm B}, J_{\rm A})$ map onto each other via the PH transformation and that, therefore, the phase diagram must be symmetric under reflection across the diagonal.
Hence, the critical points $(J_{1},0)$ and $(0,J_{1})$ map onto each other via the PH transformation, and analogously $(\infty,J_{2})$ and $(J_{2},\infty)$.

{\em A priori}, there are two different simple ways in which these four critical points can extend into the ``interior'' of the phase diagram.
Either the critical points $(J_{1},0)$ and $(0,J_{1})$ are connected by a critical line, 
and consequently, $(J_{2},\infty)$ and $(\infty,J_{2})$ by another critical line (which does not intersect the first).
This case, however, is excluded since the two sectors $\Delta N= +1$ and $\Delta N= -1$ would not be separated by a phase boundary.

Alternatively, $(J_{1},0)$ connects to $(\infty, J_{2})$ by a critical line, and consequently $(0,J_2)$ to $(J_1,\infty)$ by another critical line, which does not intersect the first and which is obtained from the first by PH transformation, i.e., by mirroring at the diagonal.
This is possible, as the two sectors $\Delta N= +1$ and $\Delta N= -1$ would indeed be separated by the $\Delta N=0$ sector.
However, this scenario is not realized either.
As Fig.\ \ref{fig:jajb} demonstrates, the critical lines do intersect at two points $(J_{\rm c1},J_{\rm c1})$ and $(J_{\rm c2}, J_{\rm c2})$ on the diagonal and merge for intermediate coupling strengths $(J_{\rm c1},J_{\rm c1}) < (J,J)<(J_{\rm c2}, J_{\rm c2})$.
As discussed above, the underlying physical mechanism is that degenerate particle-hole symmetry-broken ground states become favorable as this maximizes the gain of exchange energy.

\begin{figure}[t] 
\centering
\includegraphics[width = 0.8\linewidth]{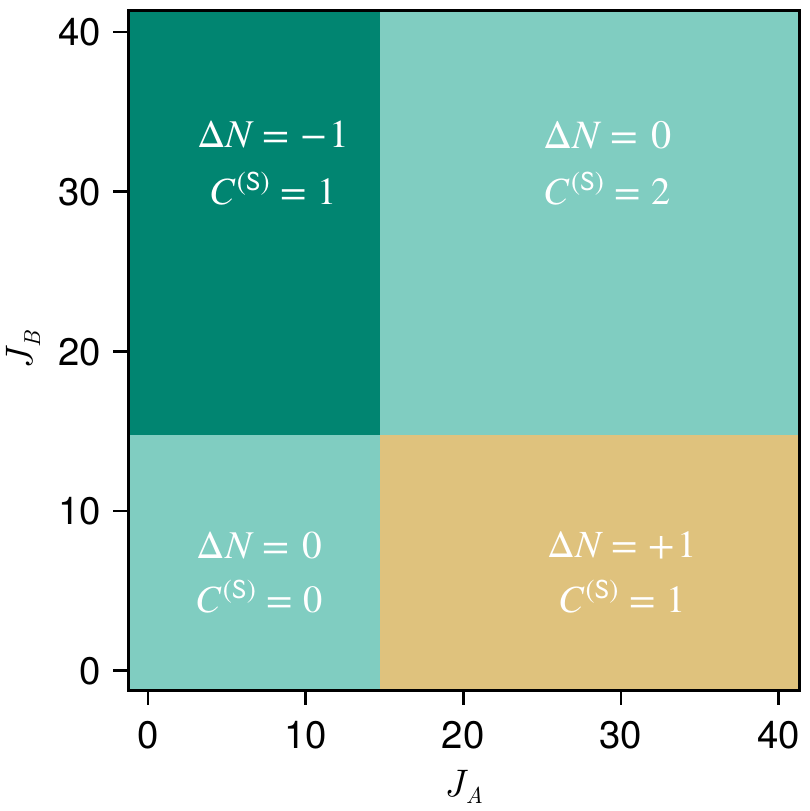}
\caption{
The same as Fig.\ \ref{fig:jajb} but for the classical-impurity-spin case.
}
\label{fig:jajbc}
\end{figure}

Finally, we briefly discuss the $J_{\rm A}$-$J_{\rm B}$ phase diagram in the classical-spin case, see Fig.\ \ref{fig:jajbc}.
S-space Chern numbers $\cs$ are easily obtained in the four limits, i.e., 
$\cs=0$ at $(J_{\rm A}, J_{\rm B}) = (0,0)$, 
$\cs=1$ at $(0,\infty)$ and at $(\infty, 0)$, 
and 
$\cs=2$ at $(\infty,\infty)$. 
Hence, we must have critical points at $(J_{1},0)$ and $(\infty, J_{2})$ along the bottom-right path, and at $(0,J_{1})$, $(J_{2},\infty)$ along the left-top path.

Along the diagonal, there is a single critical point $(J_{\rm c}, J_{\rm c})$, see Fig.\ \ref{fig:bover} and the related discussion, where the S-space Chern number changes from $\cs=0$ to $\cs=2$. 
This critical point originates as the intersection of two critical lines in the interior of the phase diagram, one that links 
$(J_{1},0)$ with $(J_{2},\infty)$
and a second one that links
$(0,J_{1})$ with $(\infty,J_{2})$.
Moving from $(0,0)$ to $(\infty, \infty)$, but slightly off the diagonal, one must thus find two critical points, consistent with the results shown in the bottom panel of Fig.\ \ref{fig:bover}.
The numerically obtained classical-spin phase diagram shown in Fig.\ \ref{fig:jajbc} confirms all these expectations. 
On the scale used for the figure, the phase boundaries appear as straight lines.

\section{Concluding discussion}
\label{sec:con}

Using a quantum many-body approach based on the Lanczos transformation and an adaptive and iterative configuration-interaction approach, we have studied generic model systems, composed of a quantum-spin impurity coupled to a gapped two-dimensional tight-binding electron-lattice model. 
The presence of a hard gap is the qualitative difference compared with the well-known Kondo problem. 
Even at zero temperature, the antiferromagnetic coupling strength $J$ must exceed a finite critical coupling $J_{\rm K}$ to reach a total-spin singlet (Kondo-screened) phase.
For $J<J_{\rm K}$, screening of the impurity spin mainly takes place via virtual second-order hopping processes by which the single-particle excitation gap $\Delta$ must be overcome, and thus the conventional Kondo effect is exponentially suppressed by the gap. 
For $J>J_{\rm K}$, screening via an additional electron in the conduction band becomes energetically favorable, i.e., $J_{\rm K}$ marks the point, at which the system's ground state switches to a sector of the Fock space with an additional electron.
Hence, the transition is characterized as a level crossing of subgap electron states and thus discontinuous.

Phase diagrams have been determined numerically for a spin-$\sfrac12$ impurity, for an $S=1$ underscreened system, and for the $S=\sfrac12$ overscreened case, where the impurity spin is coupled to both A and B orbitals at the same site $i_0$.
The underlying electron-lattice model was chosen as a prototypical Chern insulator.
We found that the dependence of $J_{\rm K}$ on the mass parameter, controlling the gap size, is qualitatively different for the topologically trivial and nontrivial phase.
One way to understand this difference relies on the presence of subgap states in the $J \to \infty$ limit, which can be understood as a remnant of the chiral gapless edge mode (after suitable deformation of the system), which is only present in the topologically nontrivial phase. 
In the impurity model and at strong $J$, it shows up as a ring state bound to the hole, created dynamically by the locally screened (or underscreened) impurity spin.
Secondly, for a large mass parameter, and thus for strong orbital polarization and large gap, a simple atomic picture is instructive to understand the phase boundaries and local correlation functions. 

Simplified variants of many-body scattering theory, which either assume an inert valence band or an inert Kondo-(under)screened state, and which thus neglect certain quantum fluctuations in a controlled way, provide a third way. 
They have turned out as very useful for understanding the $m$-dependence of $J_{\rm K}$. 
In particular, the largely different trends of $J_{\rm K}(m)$ for $S=\sfrac12$ and $S=1$ as well as the $S$-dependence of $J_{\rm K}$ in the strongly underscreened $S\to \infty$ limit could be reproduced surprisingly well.

Furthermore, we could verify the expected result that $J_{\rm K} \, (S+1)$ for $S \to \infty$ approaches the classical-spin value, i.e., the critical exchange interaction of the same system but with the quantum-spin $\hS$ replaced by a classical vector $\ff S$ of fixed length $S$, and with $J$ replaced by $\tilde{J} \equiv J \, S$.
The different scaling factors, $S+1$ vs.\ $S$, are motivated by both many-body scattering theory and the atomic limit, and lead to faster convergence.

The classical-spin and also extended multi-impurity classical-spin models are uncorrelated and thus comparatively simple systems that are exactly solvable numerically, e.g., by scattering theory, and are thus frequently employed in studies of magnetic systems coupled to gapped hosts, including superconductors.
Another insight, based on topology, is therefore useful: 
For impurity models, as studied here, topological invariants that are spatially {\em local} proved very helpful in addressing the question whether the quantum-spin and the classical-spin phase diagram are topologically equivalent, i.e., whether there are topological obstructions that prohibit a continuous deformation of these phase diagrams into each other.
Besides the k-space Chern number $\ck$, defined over the 2-torus given by the Brillouin zone, i.e., a spatially nonlocal object, we could define the B-space and the S-space Chern numbers $\cb$ and $\cs$. 
Both can take integer values only. 
The former applies to a quantum-impurity-spin model and is constructed as a Chern number $\cb$ over B-space, i.e., the 2-sphere of directions of a hypothetical local magnetic field $\ff B$ of strength $B = 0^{+}$ that couples to the impurity spin only. 
However, it is essentially useless for a classical spin. 
In the classical-spin case, we can rather define the Chern number $\cs$ over the 2-sphere of directions of the classical spin directly.
Vice versa, $\cs$ has no meaning in the quantum-spin case.

A local magnetic field $\ff B$ of {\em finite} strength has been recognized as a convenient tool to smoothly deform the quantum-spin ($B \to 0$) into the classical-spin Hamiltonian ($B \to \infty$).
Quantum-spin fluctuations are suppressed completely in the strong-$B$ limit.
For all models studied, with one exception, we observed that the continuous quantum-classical deformation of the Hamiltonian implies a continuous deformation of the corresponding phase diagrams as well.

The exception is the overscreened gapped Kondo effect, studied for the same electron system, but where a spin $S=\sfrac12$ is coupled to the electron spins in both the A and B orbitals at the same site $i_{0}$.
For equal coupling strengths $J \equiv J_{\rm A}=J_{\rm B}$, the Hamiltonian is particle-hole symmetric.
The $m$-$J$ phase diagram features an intermediate phase bounded by two critical lines, $J_{\rm c1}(m) < J < J_{\rm c2}(m)$, with two degenerate ground states in the sectors with $\Delta N = \pm 1$.
While these ground states are not invariant, they map onto each other under the PH transformation.

In the classical-spin limit, however, there is a unique, PH-symmetric ground state. 
This leads to a discontinuous quantum-spin-to-classical-spin transition, where, with increasing $B$, three quantum-spin phases, with $\cb = -1$, $\cb = 0$, and $\cb=-1$ again, must connect to only two classical-spin phases with $\cs=0$ for weak and $\cs=2$ for strong $J$.
This results in a critical point $(B_{\rm c} , J_{\rm c})$ at which the critical lines, $J_{\rm c1}(B)$ and $J_{\rm c2}(B)$, coalesce and thereafter evolve as a single critical line $J_{\rm c}(B)$. 
Beyond the critical point the spontaneously broken PH symmetry is restored. 
In the PH asymmetric model with $J_{\rm A} \ne J_{\rm B}$, the critical point vanishes, and a continuous quantum-classical deformation is regained.
The various phases in the $J_{\rm A}$-$J_{\rm B}$-$B$ parameter space are conveniently classified by local topology, i.e., by means of the B-space and S-space Chern numbers. 

We expect that the concept of local topology is powerful in the case of even more complicated systems as well. 
This includes systems with two quantum-impurity spins, where the directions of the two independent local magnetic fields form a closed four-dimensional manifold ${\sf B} \cong S^{2} \times S^{2}$. 
The topology of the bundle of ground states over ${\sf B}$, and for $B \to \infty$ over the corresponding S space ${\sf S} \cong S^{2} \times S^{2}$ formed by the directions of the classical spins, can be characterized via higher-order Chern numbers, see Ref.\ \cite{MFQ+24}.
In principle, this concept can be generalized even to Kondo-lattice-type models.
Impurity spins coupled to other electronic systems are worth studying as well. 
This includes conventional BCS and topological superconductors, beyond the one-dimensional case \cite{PP24}. 
Another class of systems, where one can profit from the comparatively simple but predictive tools of local topology to analyze phase diagrams, are impurity spins coupled to {\em correlated} electrons on a lattice, where the many-body excitation spectrum exhibits a hard gap.
Finally, further studies are needed for a better understanding of the link to the pseudogap Kondo effect for critical mass parameters at which the gap closes. 
However, this requires a different numerical approach, which is better adapted to gapless systems, e.g.\ renormalization-group techniques.

\acknowledgments
This work was supported by the Deutsche Forschungsgemeinschaft (DFG, German Research Foundation) through the research unit QUAST, FOR 5249 (project P8), project ID 449872909.
We would like to thank Maurits W. Haverkort (Heidelberg University) for discussions.

\appendix

\section{Numerical Many-Body Approach}
\label{sec:num}

For a correlated impurity coupled to a non-interacting lattice-fermion system, mapping the lattice degrees of freedom onto an effective bath is, in general, a nontrivial task. 
The Lanczos transformation \cite{BMF13} provides an efficient approach that exploits the local nature of the impurity problem while preserving the remaining symmetries of the problem. 
Moreover, it is readily applicable to multi-impurity systems as well as to lattices with only partial translational symmetry, such as those containing edges or surfaces.

Starting from one or several seed orbitals, i.e., single-particle states like $| i_{0}, A, \sigma \rangle$, the underlying single-particle lattice Hamiltonian is tridiagonalized using the (block) Lanczos algorithm \cite{GLS94}. 
By construction, the Lanczos transformation leaves the seed orbitals unchanged. 
Consequently, impurity degrees of freedom that are locally coupled to the seed orbitals remain invariant as well.

For a $D$-dimensional lattice model with finite linear extent $l$ and total number of sites $L=l^D$, the Lanczos transformation is exact (i.e., unitary) when carried out to order $d=L$. 
For an infinite system, or when truncated at order $d<l$, it generates single-particle orbitals localized within a distance of at most $d$ hopping steps from the seed orbitals. 
To control finite-size effects, the truncation order should be chosen such that, for every single-particle gap $\Delta$ under consideration, $\Delta \, d \gtrsim 1$. 
By contrast, a direct treatment of the full lattice would require the much stronger condition $\Delta \, l \gtrsim 1$, implying a total system size $L \gtrsim \Delta^{-D}$.
This substantial reduction in the required system size for a given finite-size error is a consequence of the intrinsic symmetry adaptation of the Lanczos algorithm:
Repeated action of the single-particle Hamiltonian, which respects the symmetries of the lattice, on seed orbitals, which possess their own symmetry properties, generates a Krylov subspace of states that is invariant under the transformations in the {\it intersection} of both symmetry groups. 
Consequently, only the symmetry sector relevant to the impurity problem is retained.
Nevertheless, even after this reduction, the chain lengths accessible to exact-diagonalization calculations (typically $d\sim 16$) still exhibit considerable finite-size effects.

It turns out, however, that the special structure of quantum impurity models can be exploited to reach substantially larger effective system sizes \cite{LCHH19,LHGH14}. 
The key observation is obtained from an analysis of the natural orbitals (NOs), i.e., the single-particle basis that diagonalizes the one-particle density matrix $\rho_{mm'} = \langle c_{m'}^{\dagger} c_{m} \rangle$.
Typically, only a small number of NOs are really correlated, as indicated by occupation numbers $\tilde n_i \approx \sfrac{1}{2}$. 
The occupations of the remaining orbitals approach either $0$ or $1$ exponentially, implying that correlations in these orbitals are weak and that the interaction can be treated accurately within perturbation theory.
Rotating the single-particle basis into the NO basis makes the system much more amenable to approximate wavefunction-based treatments, such as the variant of the configuration interaction (CI) approach employed here. 
To improve upon conventional CI, the present scheme first identifies an active subspace consisting of the strongly correlated NOs, defined by the criterion $| \tilde{n}_i - \sfrac{1}{2} | < \delta$ for a chosen threshold $\delta$. 
The many-body problem is solved exactly within this active space, while the remaining weakly correlated orbitals are treated through the usual CI expansion in terms of particle-hole excitations around the active-space reference state.

Grouping the natural orbitals into three classes -- the $l_c$ orbitals with occupations $\tilde n_i \approx 0$, referred to as (``conduction'') $c$-NOs, the $l_v$ orbitals with occupations $\tilde n_i \approx 1$, referred to as (``valence'') $v$-NOs, and the remaining $l_a$ strongly correlated orbitals, referred to as $a$-NOs -- a lowest-order approximation to the low-energy many-body states is obtained from the product states $|N_a,n\rangle_a \, |\rangle_c \, |\rangle_v$ .
Here, $|N_a, n \rangle_a$ denotes the $n$-th eigenstate in the $N_a$-particle sector of the subsystem spanned by the $a$-NOs, obtained by exact diagonalization within the active space. 
The states $|\rangle_c$ and $|\rangle_v$ represent the vacuum state of the $c$-NOs and the completely filled state of the $v$-NOs, respectively.

The restricted active-space configuration-interaction (RASCI) scheme partitions the many-body Hilbert space according to successive particle-hole excitations between the three orbital groups.
The reference states are characterized by particle numbers $(N_a, N_c=0, N_v=l_v)$.
Three types of particle-hole excitations are then possible, with the resulting sectors 
(i) $(N_a -1, 1, N_v)$    generated by $c^\dagger_p a_q$, with $p$, $q$ labelling respective NOs.
(ii) $(N_a+1, 0, N_v-1)$ generated by $a^\dagger_p v_q$, and 
(iii) $(N_a,     1, N_v-1)$ generated by $c^\dagger_p v_q$.
In the next order
$(N_a-2, 2, N_v)$, $(N_a-1,2,N_v-1)$, $(N_a, 2, N_v-2)$, $(N_a+1,1,N_v-2)$ and $(N_a+2,0,N_v-2)$ are possible.
Due to their cubic or higher scaling with system size, $(N_a-1,2,N_v-1)$, $(N_a, 2, N_v-2)$ and $(N_a+1,1,N_v-2)$ are neglected here.
Taking for example $(N_a-1, 1, N_v)$, the subspace is spanned by the states $\{|N_a-1, n\rangle|p\rangle_c|\rangle_v\}_{n, p}$ with $|p\rangle_c = c^\dagger_p |\rangle_c$, and the resulting matrix elements of the Hamiltonian can
easily be calculated, with the only prerequisite being the active-space eigenstate matrix elements $\langle N_a, n|a_p|N_a+1, n'\rangle$ for $N_a \in 0, \dots, (l_a-1)$.

Since the exact natural orbitals are not known {\em a priori}, the calculation must be initialized with an approximate set of orbitals, typically obtained from the corresponding non-interacting system. 
This initial guess is then refined iteratively: 
the single-particle basis is rotated into the natural-orbital basis, the ground state and the density matrix $\ff \rho$ are calculated within the RASCI scheme, and $\ff \rho$ is diagonalized to obtain an updated set of natural orbitals. 
This procedure is iterated until self-consistency is reached.

To study the system at half-filling, the chemical potential is fixed to $\mu=0$. 
The corresponding effective chain model must therefore also be solved at $\mu=0$. 
The ground state is obtained in the particle-number sector $N=N_0$ that minimizes the energy $E_0(N)$. 
Changes in the total particle number induced by level crossings are characterized by
$\Delta N(J) = N_{0}(J) - N_{0}(J=0)$, i.e., relative to the $J=0$ system.

\section{Kondo crossover at fixed $N$}
\label{sec:cross}

\begin{figure}[t] 
\centering
\includegraphics[width = 0.95\linewidth]{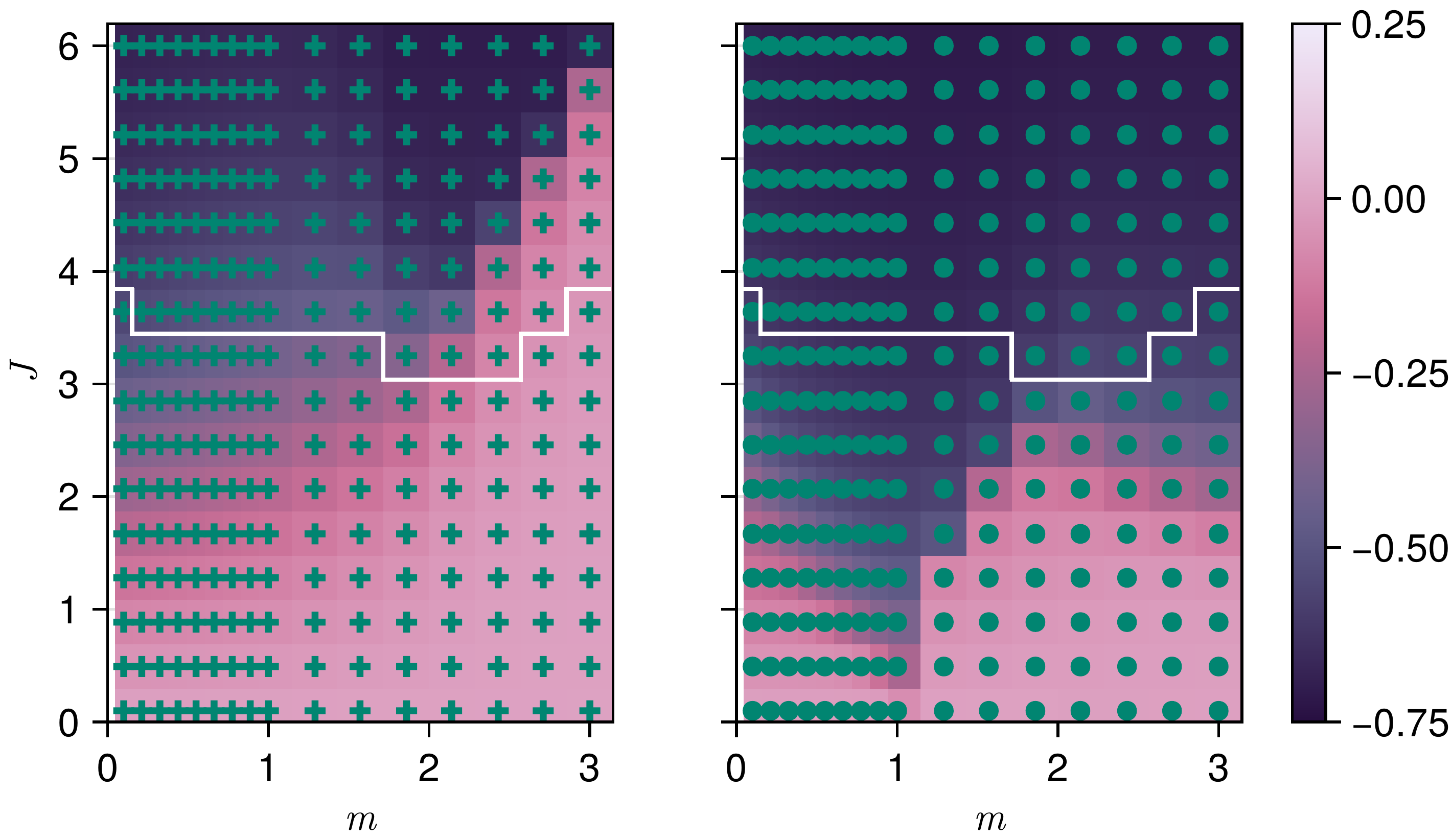}
\caption{
$m$-$J$ ground-state phase diagram as in Fig.\ \ref{fig:pd1} of the main text but for fixed particle number $N$. 
Calculations for $N = 50 + \Delta N$.
{\em Left}: $\Delta N=0$, {\em right}: $\Delta N=1$, 
{\em Background color code}: local Kondo correlation $\langle \hs \cdot \hS \rangle$. 
{\em Symbol form} indicates the ground-state degeneracy, doublet: cross, singlet: circle.
The white line indicates the discontinuous transition, given by $E_{\Delta N=0} = E_{\Delta N=1}$.
Below the line, the $\Delta N=0$ ground state is stable, above the line, the $\Delta N=1$ ground state is stable.
}
\label{fig:pdfixedn}
\end{figure}

For a gapped system, the Kondo effect is ``cut'' and the renormalization-group flow stops at an energy scale 
$T_{\rm K} \sim e^{-1/\rho_{0} J} \sim \Delta$, set by the band gap $\Delta$ \cite{Hew93}. 
This defines a critical coupling $J_{\rm K} = J_{\rm K}(\Delta)$, above which we find the Kondo-screened phase.
For $J < J_{\rm K}$, on the other hand, perturbation theory is regular and is dominated for $J \to 0$ by the linear-in-$J$ order.
This yields \cite{SGP12} a simple effective Hamiltonian $H_{\rm eff} = J \hS \cdot \hs_{\rm F}$, where $\hs_{\rm F}$ is the electron spin of the {\em singly occupied} Fermi orbital in the $J=0$ many-body Slater determinant. 
This implies that the linear-in-$J$ Kondo effect is absent in the $\Delta N=0$ sector, since at $J=0$ all single-electron states are empty or doubly occupied.

The phase diagram for fixed $\Delta N=0$ is shown in Fig.\ \ref{fig:pdfixedn} (left). 
For each value of the mass parameter $m$, we find a {\em smooth crossover} from the weak-$J$ regime with 
$\langle \hs \cdot \hS \rangle \to 0$ (unscreened impurity spin) to the strong-$J$ regime with 
$\langle \hs \cdot \hS \rangle \to - 3/4$ (local Kondo singlet).
Note that for large $m$, the crossover becomes sharper, but it remains continuous. 

In contrast, for $\Delta N=+1$, the $J=0$ ground state is constructed by filling the valence band and by adding a single electron in the conduction band occupying the Bloch state $| \ff k \sigma \rangle$ with the wave vector $\ff k = \ff k_{\rm min}$ that minimizes the gap in the BZ. 
Analysis of the band dispersions $\epsilon_{\pm}(\ff k)$, see \refeq{disp}, for the topological phase $0 < m < 2$ of the QWZ model yields that $\ff k_{\rm min} = (\pi , 0) = {\rm const.}$ for $m<1$ and $\ff k_{\rm min} = (\pi , \pi) = {\rm const.}$ for $m>1$. 
Right at $m=1$ the gap is at a maximum, and there is a flat dispersion along X--M in the BZ, such that linear superposition yields an eigenstate that is strongly localized at $i_{0}$. 
Importantly, the orbital weight discontinuously changes from $w_{\rm A}=1$ to $w_{\rm A}=0$ at $m=1$ (and vice versa for the B orbital). 
In the latter case, the above-mentioned linear-in-$J$ contribution vanishes.
Therefore, we conclude that the {\em local} spin correlation $0 > \langle \hs \cdot \hS \rangle \gtrsim  -\sfrac34$ for $m \nearrow 1$ and $J \to 0$ (but $J \ne 0$), while for $m \searrow 1$ and $J \to 0$ the impurity spin remains unscreened.
This explains the $J \to 0$, $m=1$ critical point in the $\Delta N=+1$ phase diagram, see Fig.\ \ref{fig:pdfixedn} (right).

For $\Delta N=0$, the B-space Chern number is $\cb=-1$ in the entire parameter regime, since the $\Delta N = 0$ ground state is a total spin doublet, to which $\ff B$ can couple.
For $\Delta N=1$, we have $\cb=0$, since the impurity spin is screened and a weak fictitious field cannot couple.

\section{Kondo effect for a Semenov insulator}
\label{sec:sem}

\begin{figure}[t] 
\centering
\includegraphics[width = 0.55\linewidth]{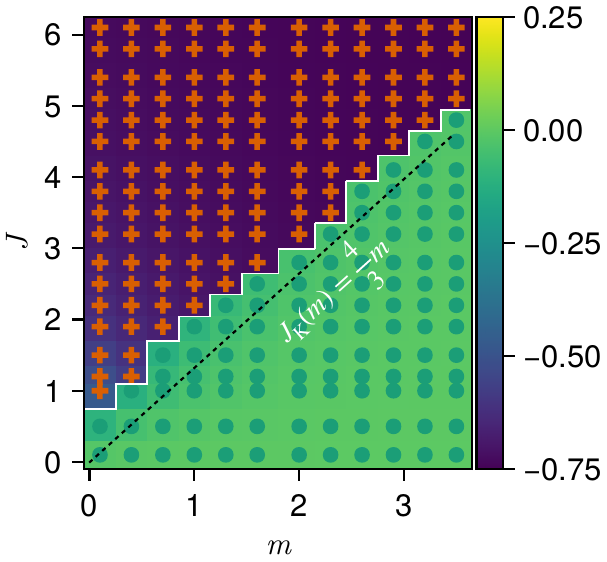}
\caption{
$m$-$J$ ground-state phase diagram as in Fig.\ \ref{fig:pd1} of the main text but for the Semenov insulator.
{\em White line}: discontinuous phase transition.
{\em Background color} code: local Kondo correlation $\langle \hs \cdot \hS \rangle \equiv \langle \hs_{i_{0} A} \hS \rangle$. 
Orange {\em symbol color}: $\Delta N = +1$, green: $\Delta N =0$. 
{\em Symbol form} indicates the ground-state degeneracy, doublet: cross, singlet: circle.
Calculations for $N = 50 + \Delta N$.
}
\label{fig:sem}
\end{figure}

We consider the Semenov-Kondo (SK) model, i.e., the two-dimensional square-lattice nearest-neighbor (n.n.) tight-binding model with a staggered on-site potential of strength $m$ \cite{Sem84}, and with a local Kondo coupling $J$ to an impurity spin $\sfrac12$ at a given site $i_{0}$:
\be
  H_{\rm SK} = \sum_{ij}^{\rm n.n.} \sum_{\sigma} t_{ij} c_{i\sigma}^{\dagger} c_{j\sigma}
  + \sum_{i\sigma} (-1)^{i} m c_{i\sigma}^{\dagger} c_{i\sigma}
  + J \hS \cdot \hs_{i_{0}}
\: . 
\labeq{semham}  
\ee
The chemical potential is set to $\mu=0$ such that the system is half-filled.
For $m=0$, the system is a noninteracting metal, and any $J>0$ will lead to a screening of the impurity spin (at zero temperature). 
A finite $m$ opens a band gap and thus leads to a gapped Kondo effect.
The formation of a singlet, i.e., the screening of the impurity spin $\hS$, sets in at a finite Kondo coupling $J_{\rm K}(m) > 0$. 
In the atomic limit $t_{ij}=0$, basically the same reasoning as for the QWZ model, yields $J_{\rm K}(m) = \sfrac43 \, m$. 
Since the Semenov insulator is a trivial (non-topological) band insulator for $m>0$, we do not expect a substantial effect of screening via virtual processes overcoming the band gap $2m$.
In fact, the phase diagram, as obtained numerically, see Fig.\ \ref{fig:sem}, shows $\langle \hs \cdot \hS \rangle \approx 0$ for $J< J_{\rm K}(m)$, except for smaller $m$, due to proximity to the metallic phase.
Note that the limit $m\to 0$ cannot be accessed reliably, given the practical limitations on the system size and total electron number $N$. 
Computations have been performed for $N = 50 + \Delta N$ with $\Delta N =0,1,2$. 
Figure \ref{fig:sem} demonstrates that the rough atomic-limit estimate for $J_{\rm K}(m)$ nicely describes the main trend.

%


\end{document}